\documentclass[12pt]{iopart}

\usepackage[square,numbers,sort&compress]{natbib}
\usepackage{iopams} 
\usepackage{algorithm}
\usepackage{algpseudocode}
\usepackage{tikz}
\usepackage{breqn}
\usepackage{etoolbox}

\newrobustcmd*{\myVtriangle}[2]{\tikz{\filldraw[draw=#1,fill=#2] (0cm,0.2cm) --
(0.2cm,0.2cm) -- (0.1cm,0cm) -- (0cm,0.2cm);}}

\newrobustcmd*{\mythickVtriangle}[2]{\tikz{\filldraw[line width=0.3mm,draw=#1,fill=#2] (0cm,0.2cm) --
(0.2cm,0.2cm) -- (0.1cm,0cm) -- (0cm,0.2cm);}}

\newrobustcmd*{\mytriangle}[2]{\tikz{\filldraw[draw=#1,fill=#2] (0.0cm,0.0cm) --
(0.2cm,0cm) -- (0.1cm,0.2cm) -- (0cm,0cm);}}

\newrobustcmd*{\mysquare}[2]{\tikz{\draw[draw=#1,fill=#2] (0cm,0cm)
rectangle (0.2cm,0.2cm)}}

\newrobustcmd*{\mythicktriangle}[2]{\tikz{\filldraw[line width=0.3mm,draw=#1,fill=#2] (0.0cm,0cm) --
(0.2cm,0cm) -- (0.1cm,0.2cm) -- (0.0cm,0cm);}}

\newrobustcmd*{\mythicksquare}[2]{\tikz{\draw[line width=0.3mm,draw=#1,fill=#2] (0cm,0cm)
rectangle (0.2cm,0.2cm)}}

\newrobustcmd*{\mybarredtriangle}[2]{\tikz{\draw[draw=#1,fill=#2] (0,0) --
(0.2cm,0) -- (0.1cm,0.2cm) -- (0cm,0cm); \draw[draw=#1] (-0.1cm, 0.07cm) -- (0.3cm, 0.07cm)}}

\newrobustcmd*{\mythickbarredtriangle}[2]{\tikz{\draw[line width=0.3mm,draw=#1,fill=#2] (0,0) --
(0.2cm,0) -- (0.1cm,0.2cm) -- (0cm,0cm); \draw[draw=#1] (-0.1cm, 0.07cm) -- (0.3cm, 0.07cm)}}

\newrobustcmd*{\mybarredsquare}[2]{\tikz{\draw[draw=#1,fill=#2] (0,0)
rectangle (0.2cm,0.2cm); \draw[draw=#1] (-0.1cm, 0.1cm) -- (0.3cm, 0.1cm)}}

\newrobustcmd*{\mythickbarredsquare}[2]{\tikz{\draw[line width=0.3mm,draw=#1,fill=#2] (0,0)
rectangle (0.2cm,0.2cm); \draw[draw=#1] (-0.1cm, 0.1cm) -- (0.3cm, 0.1cm)}}

\newrobustcmd*{\mybarredcircle}[2]{\tikz{\draw[draw=#1,fill=#2] (0,0)
circle (0.1cm); \draw[draw=#1] (-0.2cm, 0.0cm) -- (0.2cm, 0.0cm)}}

\newrobustcmd*{\mythickbarredcircle}[2]{\tikz{\draw[line width=0.3mm,draw=#1,fill=#2] (0,0)
circle (0.1cm); \draw[draw=#1] (-0.2cm, 0.0cm) -- (0.2cm, 0.0cm)}}

\newrobustcmd*{\mydashedline}[1]{\tikz{\draw[draw=#1] (-0.2cm, 0.2cm) -- (-0.1cm, 0.2cm); \draw[draw=#1] (-0.0cm, 0.2cm) -- (0.1cm, 0.2cm)}}

\newrobustcmd*{\mythickcross}[1]{\tikz{\draw[line width=0.3mm,draw=#1] (0,0) --
(0.2cm,0); \draw[line width=0.3mm,draw=#1] (0.1cm,-0.1cm) -- (0.1cm,0.1cm);}}

\newrobustcmd*{\mybarredcross}[1]{\tikz{\draw[line width=0.3mm,draw=#1] (0,0) --
(0.2cm,0); \draw[line width=0.3mm,draw=#1] (0.1cm,-0.1cm) -- (0.1cm,0.1cm); \draw[draw=#1] (-0.1cm,0) -- (0.3cm,0);}}

\newrobustcmd*{\myline}[1]{\tikz{\draw[draw=#1] (-0.15cm, 0.1cm) -- (0.15cm, 0.1cm);\draw[line width=0.3mm,draw=#1] (-0.0cm, 0.0cm);}}

\newrobustcmd*{\mythickline}[1]{\tikz{\draw[line width=0.3mm,draw=#1] (-0.15cm, 0.1cm) -- (0.15cm, 0.1cm);\draw[line width=0.3mm,draw=#1] (-0.0cm, 0.0cm);}}

\newrobustcmd*{\mythickdashedline}[1]{\tikz{\draw[line width=0.3mm,draw=#1] (-0.2, 0.1cm) -- (-0.1cm, 0.1cm); \draw[line width=0.3mm,draw=#1] (-0.0cm, 0.1cm) -- (0.1cm, 0.1cm); \draw[line width=0.3mm,draw=#1] (-0.0cm, 0.0cm);}}

\newrobustcmd*{\mythickdasheddottedline}[1]{\tikz{\draw[line width=0.3mm,draw=#1] (-0.22, 0.1cm) -- (-0.13cm, 0.1cm); \draw[line width=0.3mm,draw=#1] (-0.085cm, 0.1cm) -- (-0.055cm, 0.1cm); \draw[line width=0.3mm,draw=#1] (-0.01cm, 0.1cm) -- (0.08cm, 0.1cm); \draw[line width=0.3mm,draw=#1] (-0.0cm, 0.0cm);}}

\newrobustcmd*{\mycircle}[2]{\tikz{\draw[draw=#1,fill=#2] (0,0)
circle (0.1cm);}}

\newrobustcmd*{\mythickcircle}[2]{\tikz{\draw[line width=0.3mm,draw=#1,fill=#2] (0,0)
circle (0.1cm);}}

\newrobustcmd*{\mydot}[1]{\tikz{\draw[line width=0.3mm,draw=#1] (0,0)
circle (0.025cm);}}

\begin{document}

\newcommand{\alert}[1]{{\color{red} #1}}
\newcommand{\bbxi}{\boldsymbol{\xi}}
\newcommand{\bg}{\boldsymbol{\Gamma}}
\newcommand{\bgsv}{\boldsymbol{g}}
\newcommand{\bx}{\boldsymbol{x}}
\newcommand{\bk}{\boldsymbol{\kappa}}
\newcommand{\bbphi}{\boldsymbol{\phi}}

\newcommand{\norm}[1]{\left\lVert#1\right\rVert}

\title{A self-similarity principle for the computation of rare event probability}

\author{Malik Hassanaly and Venkat Raman}

\address{Department of Aerospace Engineering, University of Michigan, Ann Arbor, MI}
\ead{malik.hassanaly@gmail.com}
\vspace{10pt}
\begin{indented}
\item[]Oct 2019
\end{indented}

\begin{abstract}
The probability of rare and extreme events is an important quantity for design purposes. However, computing the probability of rare events can be expensive because only a few events, if any, can be observed. To this end, it is necessary to accelerate the observation of rare events using methods such as the importance splitting technique, which is the main focus here. In this work, it is shown how a genealogical importance splitting technique can be made more efficient if one knows how the rare event occurs in terms of the mean path followed by the observables. Using Monte Carlo simulations, it is shown that one can estimate this path using less rare paths. A self-similarity model is formulated and tested using an a priori and a posteriori analysis. The self-similarity principle is also tested on more complex systems including a turbulent combustion problem with $10^7$ degrees of freedom. While the self-similarity model is shown to not be strictly valid in general, it can still provide a good approximation of the rare mean paths and is a promising route for obtaining the statistics of rare events in chaotic high-dimensional systems.
\end{abstract}

%\maketitle

\section{Introduction}
Rare excursions of a complex system from its nominal behavior are difficult to predict, since very few observations, if at all any, would exist to reliably characterize the source of this behavior. Although rare events are sparsely encountered, when they are associated with extreme events, their impact can be of tremendous importance. For instance, in propulsion applications, gas turbines and other such energy conversion devices are designed to minimize failure probability, but any failure can have catastrophic consequences \cite{hassanaly2019computational}. The problem of rare event prediction is also relevant to other fields such as market crashes in financial systems \cite{sornette1996stock}, prediction of reaction rates in molecular dynamics \cite{weinan2007simplified}, the occurrence of rogue waves next to offshore platforms \cite{solli2007optical}, rare and extreme atmospheric heat waves \cite{ragone2018computation}, heavy rains \cite{ghil2011extreme,frei2001detection} or energy grid blackouts \cite{kim2013splitting}. Even though each such event is rare, the presence of this possibility requires an endeavor to develop prediction tools. 

Defining prediction in this context is itself a multifaceted issue \cite{hassanaly2019computational}, but much of the prior work in this area can be grouped as following a dynamical systems approach or a statistical approach. In the dynamical approach, the goal is to characterize the dynamical behavior of the system, such as its response to perturbations or its stability properties \cite{hassanaly2018ensemble}. One application could be that understanding these aspects could enable real-time control, where some precursor is identified for the purpose of an actuation mechanism. In the statistical approach, the goal is to obtain statistics of a rare event. Estimating the probability of a rare event may enable a more resilient system that is less susceptible to such extreme excursions. Such statistical approaches can also reveal the average behavior close to a rare event, which may also be used to identify precursors. However, developing such a statistically significant ensemble of rare events is in itself a formidable challenge. This is especially the case in high-dimensional systems, where several paths to such rare events may exist. The focus of this work is to develop a statistical framework for such high-dimensional systems, where features of a rare event trajectory are inferred from less rare (more probable) trajectories. 

Rare events occur when there is some uncertainty imposed on the system. For instance, this could be due to uncertainty associated with the initial or boundary conditions. The objective of the numerical approach is to sample realizations that lead to extreme events, as defined by specific regions of phase-space. For this purpose, consider the quantities used to define the rare events as $\bbxi \in E$, where $E$ is the phase space. The rare event probability is defined as $p = P(\bbxi \in A)$, where $A \subset E$. In a Monte-Carlo sense, this probability can be estimated by sampling $n$ independent events, with the estimator of $p$ given by
\[
p = \frac{1}{n} \sum_{i=1}^n 1_A(\bbxi_i),
\]
where $1_A(x) = 1$ if $x \in A$ and $0$ otherwise. While the estimator is unbiased, the variance of this estimator can be obtained as $(p-p^2)/{n}$. The relative uncertainty of the estimator scales as $\sqrt{\frac{1}{np}}$ for $p \ll 1$. Therefore, the lower the probability to be estimated, the higher relative uncertainty of the estimator. In order to improve this estimation process, a reliable approach to sampling more of the trajectories that lie in $A$ is necessary. The main challenge that makes such an estimator difficult to obtain is that the set of initial/boundary conditions that lead to the rare event is not known a priori. 

There are many techniques that exist to improve this estimation procedure. For example, in importance sampling methods \cite{siegmund1976importance}, a biased distribution is used to draw more samples from certain regions of initial and boundary condition space. However, the optimal bias needed to obtain extreme events is unknown, which reduces the effectiveness of this method \cite{morio2014survey}. Typically, a functional form of the biased distribution is assumed, and its parameters are adjusted to approximate the optimal biased distribution \cite{de2005tutorial}. The biased distribution could also be obtained from the direct observation of a rare event with a cheaper low-fidelity model \cite{peherstorfer2016multifidelity}. An alternative technique is the importance splitting (ISP) method, where the rare event region is progressively reached. Suppose that one defines a rare event as $A = \{  \bbxi, Q(\bbxi) > a \} $, where $\bbxi$ is the state of the system, $Q(\bbxi) \in \mathbb{R}$ is the quantity of interest (QoI) and $a$ is some threshold. Instead of directly finding $A$, one can sample $A_i = \{  \bbxi, Q(\bbxi) > a_i \}_{1 \leq i \leq N}$ such that $a_1 < ... < a_N$, and use the fact that $A_1 \supset ... \supset A_N$. Until recently, these methods were limited due to their sensitivity to the definition of the intermediate levels $a_1, ..., a_N$. New adaptive methods that do not require to define these levels explicitly \cite{cerou} have allowed importance splitting to be used in a variety of fields \cite{bouchet2019rare,teo2016adaptive}. To define the levels, multiple realizations are run from a baseline state. The state obtained that is the closest to the rare state is then selected as a starting point for the next generation of realizations. In this framework, simulations are spawned at non-regular intervals and the rare event can occur after an arbitrarily long time. Recently, such techniques have been shown to be particularly successful at capturing rare events in turbulent flows. For example, it has been used to observe transitions in planetary atmosphere \cite{laurie2015computation,bouchet2019rare}. A similar algorithm, close to the one investigated here, uses realizations that are killed or cloned at regular intervals depending on their estimated potential of reaching a rare event. There, the rare event can be defined at a pre-specified time. This technique has also recently been successful for computing the time separating two extreme events (return times) in turbulent flows that can be obtained from the probability of the rare event itself \cite{ragone2018computation,lestang2018computing}

For the specific problems of interest, Wouters et al.~\cite{wouters_bouchet} analyzed a genealogical particle algorithm that was introduced in Ref.~\cite{del2005genealogical}. The algorithm relies on the simultaneous evolution of several copies of the dynamical system. The copies that are the closest to the rare event are cloned or pruned periodically. At the end of the procedure, the observations of the system are clustered next to rare event and can be used to estimate the desired probability. Wouters et al.~\cite{wouters_bouchet} also recognized that the ISP algorithm could be improved with prior knowledge of the trajectory leading to a rare event. In particular, the approach is inspired by techniques used in computational chemistry. For instance, if the goal is to capture the probability of transition of a molecular system from one state to another, it can be beneficial to focus calculations around the most likely transition path in order to observe many such transitions \cite{dellago2002transition}. In other areas, including fluid mechanics, the importance of the most likely path leading to a rare event (the instanton) has also been recognized and used to sample rare events \cite{zhang2018rare,laurie2015computation,grafke2015instanton}. However, there is no known robust method to compute the path to rare events in deterministic chaotic system.
 
 With this background, the paper proposes a simple method to estimate the trajectory leading to a rare event in a deterministic and chaotic system. This trajectory is then used with the ISP algorithm studied by Wouters et al.~\cite{wouters_bouchet} to measure its effect on the computational gain. The computational gain is computed in the case of high-dimensional deterministic dynamics, relevant for turbulent flow applications.
 
 The rest of the paper is organized as follows: Section~\ref{sec:rarePath} briefly describes the algorithm analyzed by Wouters et al. and illustrates the benefit of using the average time-history of the observable that correspond to rare events for the statistics of the probability estimator. Section~\ref{sec:estimateRarePath} introduces a method to estimate the path to a rare event from a simple rationale. The method is tested with an \textit{a priori} and \textit{a posteriori} analysis. In Sec.~\ref{sec:applicability}, the applicability of the method to more complex systems is investigated and discussed. Concluding remarks are provided in Sec.~\ref{sec:conclusions}.

\section{Path-to-event based estimation of probability}
\label{sec:rarePath}

\subsection{Definition of the dynamical system and rare event notation}

The rare events of interest occur over a finite-time interval, and the quantity of interest (QoI) is defined at the final time. More formally, the dynamical system is defined as 
\begin{equation}
    \forall t \in [0, T_f], ~\frac{d\bbxi}{dt} = \mathcal{F}(\bbxi),~\bbxi(t=0) \sim \mathcal{P},
\end{equation}
where $\mathcal{F}$ represents the governing equations. In the case of turbulent flow or other systems governed by partial differential equations, ${\cal F}$ is some finite-dimensional approximation of these equations obtained by, for instance, numerical discretization. $\mathcal{P}$ is the nominal distribution of initial conditions. The QoI is defined as $QoI=Q(\bbxi(T_f))=Q(T_f)$ and the observable is chosen to be $Q(t)$. In the rest of the document, Q is chosen to be a single scalar.

\subsection{Genealogical particle algorithm}
\label{sec:algo}

The genealogical particle algorithm studied in Ref.~\cite{wouters_bouchet} is briefly presented as it will be used as the basis for the proposed method. Since the problem is deterministic, only the uncertainty in initial conditions can lead to a distribution of QoI. The goal is to find the probability with which the QoI exceeds a certain threshold $a$, $P(Q(T_f) > a)$. The algorithm starts by sampling $M$ initial conditions (also called particles) at $t=0$ that are evolved over time with timestep $\Delta t$. The name genealogical is due to the fact that at intermediate times (or selection steps) between $t=0$ and $t=T_f = N_t \Delta t$, realizations that are deemed likely to lead to the rare event are cloned, while others are pruned. Cloning of particles implies that at time $t=T$, another copy of this particle is initiated. For deterministic dynamics, this copy should be perturbed by a small amount before being evolved. In this work, the selection steps are fixed a priori and occur at fixed time stations. For the numerical tests that are conducted, the number of selection steps is expressed in terms of the number of timesteps $N_s$ that separate two selections. After every selection step, a special procedure is used to maintain the total number of trajectories to be constant (see \cite[Sec. 2.3.3]{wouters_bouchet} for the details of this procedure). In order to select the particles that need to be cloned or pruned at every selection step, a weight is attributed to each particle based on the chosen observable. For additional clarity, the algorithm described above is summarized as pseudo code in Algo.~\ref{algo:wouters}. Step 6 and 12 of Algo.~\ref{algo:wouters} are explained in detail in Ref.~\cite{wouters_bouchet} and are not expanded further. The focus of the present work is on Step 5.

\begin{algorithm}
\caption{Genealogical Importance splitting algorithm}\label{algo:wouters}
\begin{algorithmic}[1]

\State Sample $M$ initial conditions ($t=0$)
\State j=0
\For{$i=1,N_t$}
\If{$j==N_s$}
\State{Compute the weight $W$ of each sample.}
\State{Prune samples with low weight and clone samples with high weight.}
\State{j=0}
\EndIf
\State{Evolve samples over $\Delta t$.}
\State{j=j+1}
\EndFor
\State{Compute probability estimator}
\end{algorithmic}
\end{algorithm}

The weight chosen in Ref.~\cite{wouters_bouchet} is of the form $W=exp(C \Delta Q)$, where $C$ is a constant called the \textit{weighting factor} and $\Delta Q$ is the variation of the observable between two selection steps. The value of $C$ determines the aggressiveness of the cloning process expressed as the pruning ratio $\frac{N_{prune}}{M}$, where $N_{prune}$ is the number of particles that are pruned and M is the total number of particles evolved. A large value of $C$ would lead to a large pruning ratio, meaning that at every selection step, only a few particles are saved. In turn, a large pruning ratio would also prioritize particles that lead to the rarest events. In Ref.~\cite{wouters_bouchet}, it was proposed to choose $C$ such that $C = \frac{\Delta \mu_Q}{\sigma_Q^2}$, where $\Delta \mu_Q=E_{ISP}(Q(T_f)) -  E(Q(T_f))$, the difference between the desired average of the QoI at $T_f$ over the $M$ particles used in the ISP and the unbiaised expected value of $Q(T_f)$, i.e. when no selection is applied; $\sigma_Q^2 = Var(Q(T_f))$ is the unbaised variance of $Q$ at $t=T_f$, i.e. when no selection is applied. In practice, the statistics of the QoI can be estimated by running $M$ non-biased simulations prior to using the ISP. If a large deviation from the average behavior is sought, then $\Delta \mu_Q$ will be larger and the value of $C$ will increase as well. Since $C$ controls the pruning process, it is necessary to understand the implications of the choice of this variable. 

\begin{figure}[h]
\begin{center}
\includegraphics[width=0.3\textwidth]{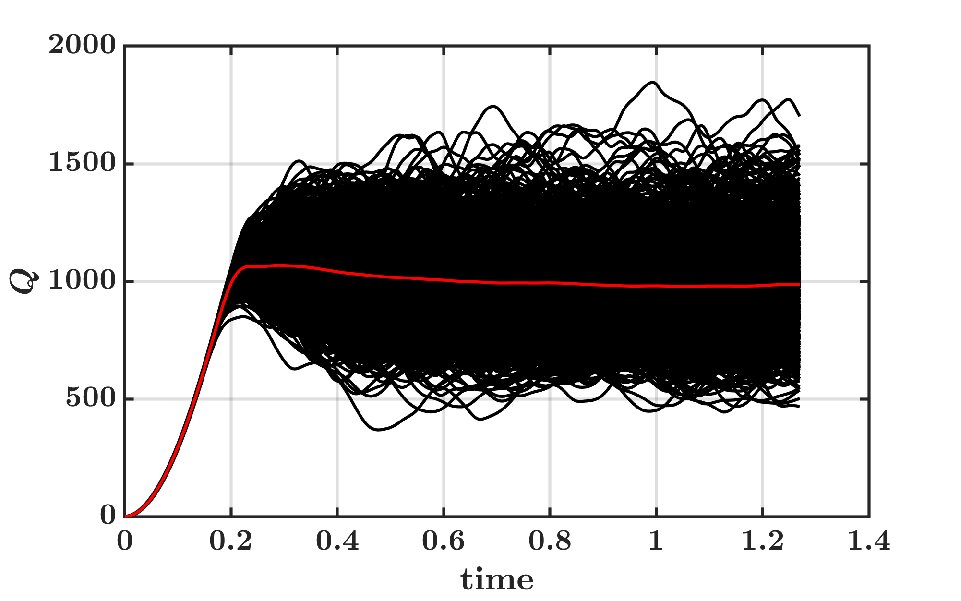}
\includegraphics[width=0.3\textwidth]{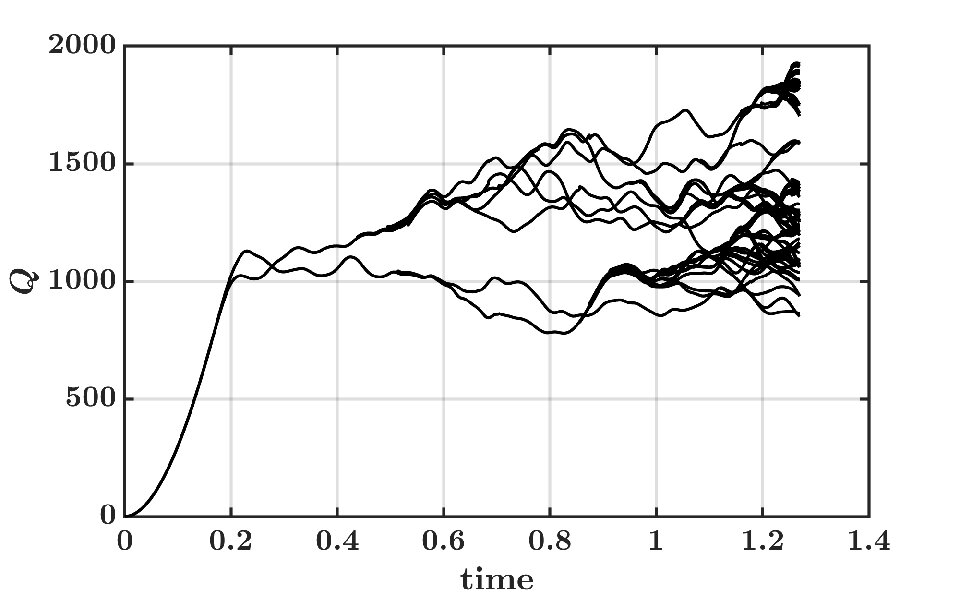}
\includegraphics[width=0.3\textwidth]{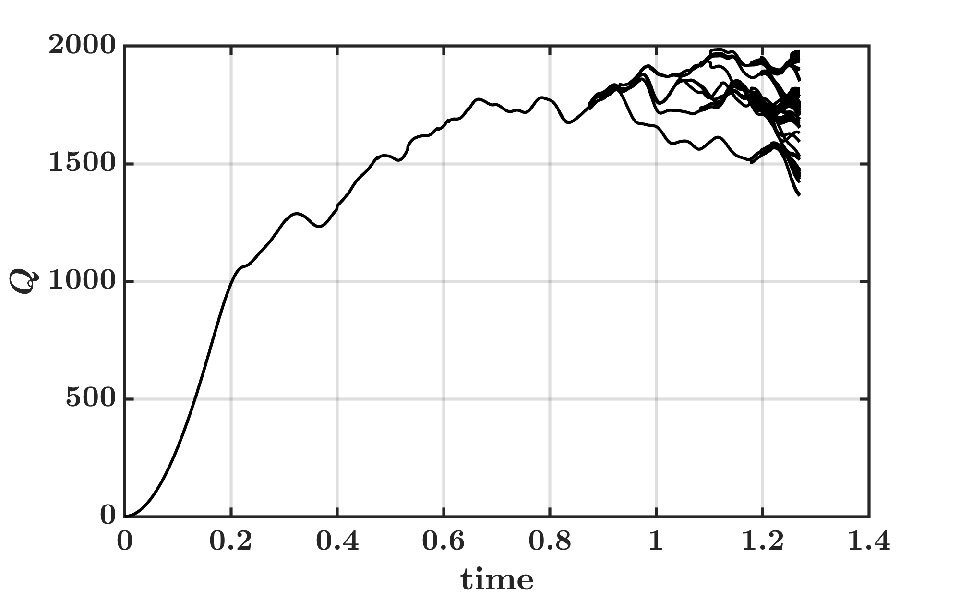}
\caption{Left: instantaneous values of the observable $Q(t)$ plotted over time (\mythickline{black}). Ensemble average of $Q$ (\mythickline{red}). Middle: time-history of $Q(t)$ obtained with the importance splitting algorithm using a constant weight factor $C=0.0104$. Right: time-history of $Q$ obtained with the importance splitting algorithm using a constant weight factor $C=0.0208$. Plots correspond to the Lorenz 96 case with 32 degrees of freedom.}
\label{fig:illL96}
\end{center}
\end{figure}

The influence of the weighting factor $C$ on the ISP is illustrated in Fig.~\ref{fig:illL96} for the Lorenz 96 system \cite{lorenz1996predictability}, using parameters based on Ref.~\cite{wouters_bouchet}. This system will be used throughout the paper as a benchmark for the performances of the modified algorithm. This system is non-linear, can have an arbitrarily high number of dimension and exhibits chaotic dynamics that makes it a good surrogate for turbulent flow problems. The governing equations are written as

\begin{equation}
    \forall t \in [0,T_f], ~\forall i \in [1,d], \frac{d\xi_i}{dt} = \xi_{i-1} (\xi_{i+1} - \xi_{i-2}) + f - \xi_i,
    \label{eq:l96}
\end{equation}

where $f=256$, $\xi_{d+1}=\xi_1$,  $\xi_{0}=\xi_d$, $\xi_{-1}=\xi_{d-1}$ and $d$ is the number of degrees of freedom chosen (here, 32). The observable considered here is related to the turbulent energy of the system and is defined by
\begin{equation}
    Q=\frac{1}{2 d} \sum_{i=1}^d \xi_i^2.
    \label{eq:qoi}
\end{equation}

A second-order Runge Kutta scheme is used for the time-integration, and the time step is set to $10^{-3}$. In Fig.~\ref{fig:illL96} (left), 1000 instantaneous realizations of $Q$ are shown along with the ensemble average. The initial conditions are normally distributed, i.e.:
\begin{equation}
    \xi_i(t=0) \sim \mathcal{N}(0, 1), \forall i
    \label{eq:ic}
\end{equation}
and quickly diverge from each other due to the chaotic dynamics. In the middle and right plots, the trajectories simulated at the end of the splitting algorithm using $M=100$ particles are shown for weighting factors $C=0.0104$ (middle) and $C=0.0208$ (right). Note that the weighting factor $C=0.0104$ was used in Ref.~\cite{wouters_bouchet} to demonstrate the capabilities of the algorithm. At every selection step, the clones are perturbed using a normal distribution $\mathcal{N}(0,0.871)$. The selection steps are separated by $N_s=19$ timesteps. The magnitude of the perturbations should be large enough to generate realizations that can be considered independent at the next selection time, and small enough to not affect the probability computed. Similarly, selection frequency should be low enough to ensure that realizations are not too correlated, but large enough so that realization do not have time to return to non-rare paths. These are hyperparameters of the method. In Ref.~\cite{wouters_bouchet}, a sensitivity analysis is used to set them. Acceptable values found there are used in the present study.

The importance splitting algorithm is stochastic due to perturbation introduced during the cloning process. Hence, the algorithm needs to be executed multiple times to obtain statistics about the probability estimator. As can be seen, the largest weighting factor leads to the largest value of QoI at the final time. However, it also quickly discards most of the particles. This effect was also observed in \cite{nemoto2016population}. As a result, the solution space ensemble containing the trajectories that lead to a rare events are not appropriately sampled, which can lead to large variance for the probability estimator, thereby increasing its associated uncertainty.

\subsection{The rare mean path approach}
\label{sec:fluc}

In order to improve the trajectory sampling for estimating probabilities of events significantly removed from the mean behavior, the weighting factor will have to be modified. In fact, the weighting factor could be made time-dependent \cite{wouters_bouchet}:
\begin{equation}
    C(t) = \frac{\mu_{rare}(t) - \mu_{Q}(t)}{\sigma_Q(t)^2},
    \label{eq:weight}
\end{equation}
where $\mu_{rare}(t) = E_{rare}(Q(t))$ is the average $Q(t)$ conditioned on all the trajectories that lead to a rare event, $\mu_{Q}(t) = E(Q(t))$ is the time-dependent average of $Q(t)$ and $\sigma_Q(t)^2$ is the time-dependent variance of $Q(t)$. In Ref.~\cite{wouters_bouchet}, this method was illustrated for a noise-driven problem for which $C(t)$ could be analytically computed, and it was shown to decrease the pruning ratio while leading to the expected final QoI. Here, the objective is to evaluate this procedure for the Lorenz 96 problem and quantify the impact of the procedure on the statistics of the probability estimator. In all the following discussions, the name \textit{target level} refers to the threshold that QoI is designed to exceed at the final time when using the ISP. For the fixed-weight algorithm, target level refers to $E_{ISP}(Q(T_f))$ and for the time-dependent weight algorithm, target level refers directly to the threshold $a$. 

The system of interest is defined by Eq.~\ref{eq:l96}. While, $\sigma_Q(t)$ and $\mu_{Q}(t)$ can be approximated using a few observations of the system, estimating $\mu_{rare}(t)$ requires observation of the rare event. The target levels investigated are $Q(T_f)=1356$ and $Q(T_f)=1737$, and are located two and four standard deviations away from the mean. In order to get a reasonable approximation of the mean path leading to these values, $10^6$ trajectories are simulated. The value of the observables for the trajectories exceeding these levels are stored and averaged at every timestep. More formally, the mean path $R(t)$ that exceeds a threshold $a$ is defined as:

\begin{equation}
    \forall t \in [0, T_f], ~R_a(t) = \langle~ Q(t)~|~Q(T_f) > a~ \rangle.
    \label{eq:defRareMeanPath}
\end{equation}

\begin{figure}[h]
\begin{center}
\includegraphics[width=0.45\textwidth]{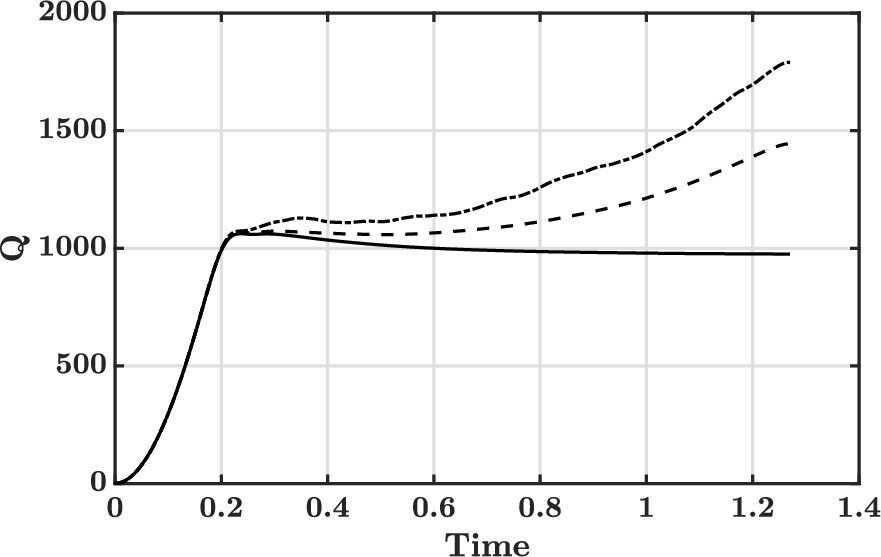}
\caption{Ensemble average time history of the observable (\mythickline{black}). Rare mean path of the observable $R_{1356}$ (\mythickdashedline{black}); $R_{1737}$ (\mythickdasheddottedline{black}). Plots correspond to the Lorenz 96 case with 32 degrees of freedom.}
\label{fig:meanRarePath}
\end{center}
\end{figure}

The mean trajectories of the observable leading to both levels are shown in Fig.~\ref{fig:meanRarePath}.

\begin{figure}[h]
\begin{center}
\includegraphics[width=0.32\textwidth]{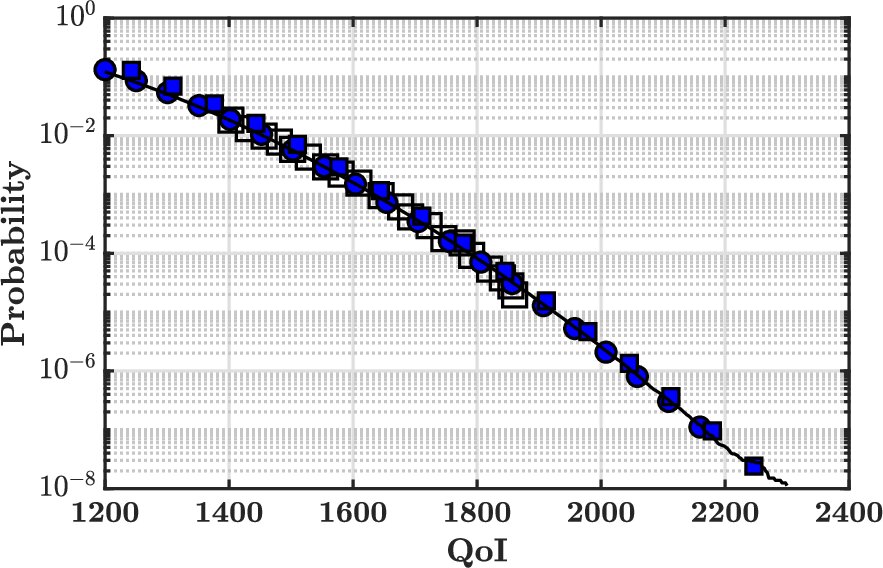}
\includegraphics[width=0.32\textwidth]{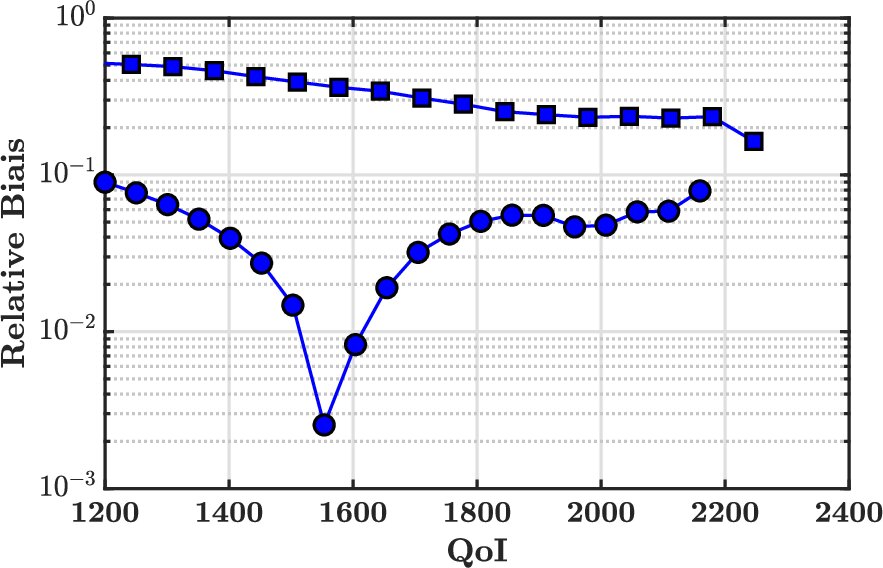}
\includegraphics[width=0.32\textwidth]{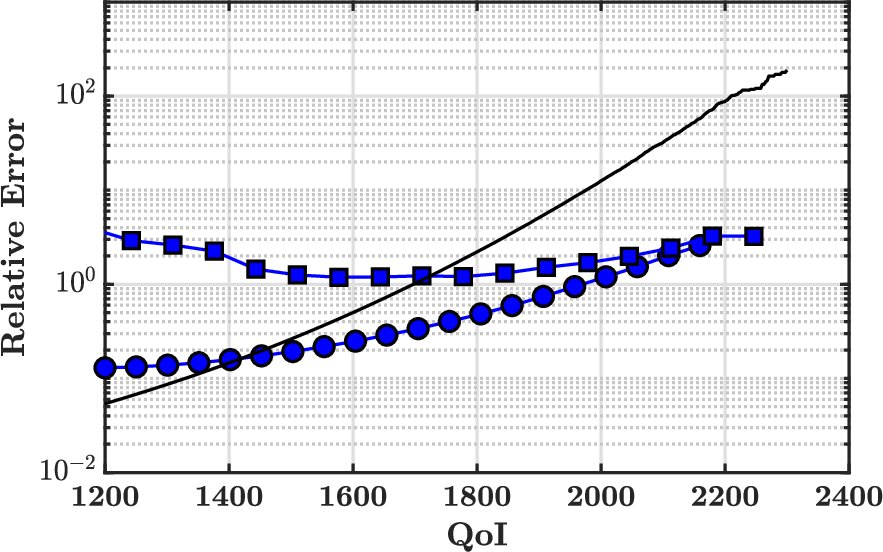}
\caption{In all the plots the ISP method with time-dependent weight based on $R_{1356}$ (\mythickcircle{black}{blue}), and $R_{1737}$ (\mythicksquare{black}{blue}) and $M=2500$ particles are denoted by the same symbols. Left: complementary of the cumulative density function (CDF) of the QoI. Probabilities are compared to the one obtained with brute force calculation (\mythickline{black}). Middle: biais of probability estimator normalized by the probability obtained from brute force calculation. Right: relative error of the estimators compared with the relative error of a brute force calculation that would use $M=2500$ samples (\mythickline{black}) calculated as $\sqrt{\frac{p-p^2}{M p^2}}$, where $p$ is the probability computed from the brute force calculation with $8 \times 10^8$ samples. Plots correspond to the Lorenz 96 case with 32 degrees of freedom.}
\label{fig:probL96_32}
\end{center}
\end{figure}

Using the computed rare mean path, the probability $P(Q(T_f)>1356)$ and $P(Q(T_f)>1737)$ are computed using the genealogical particle algorithm presented in Sec.~\ref{sec:algo} with the fixed and the time-dependent weighting factor. Here, the number of particles is $M=2500$ and algorithm is run 100,000 times in order to ensure that the statistics of the estimator are converged. In Fig.~\ref{fig:probL96_32}, the probability obtained with the fluctuating path approach is compared to a brute force calculation computed with $8 \times 10^8$ calculations. Note that although a single level is targeted, it is only used to compute $C$. The ISP method used does not preclude from estimating probabilities at higher levels. Therefore, the results present probabilities corresponding to different value of the QoI. As can be seen the variance is significantly reduced compared to that of the Monte-Carlo simulation at the levels targeted. The biais is not exactly equal to zero despite repeating the simulations 100,000 times which could be attributed to the cloning method. Nevertheless, the relative biais at low probabilities is much lower than the relative error, and the ISP estimator can be considered unbiaised. Furthermore, by targeting larger levels, lower probabilities can be estimated by the algorithm. For reference, the probability obtained in Ref.~\cite{wouters_bouchet} through Monte-Carlo run are also indicated to show consistency between the cases run. 

\begin{figure}[h]
\begin{center}
\includegraphics[width=0.45\textwidth]{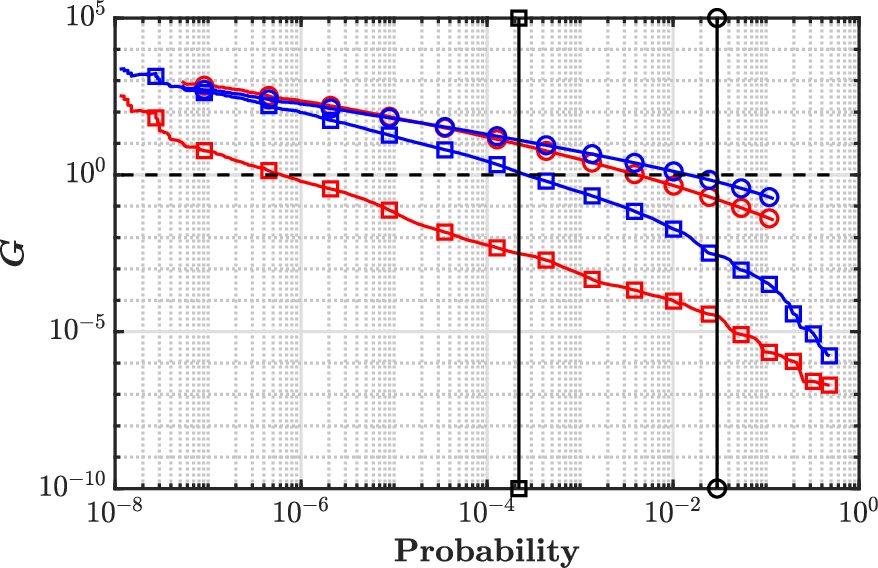}
\includegraphics[width=0.45\textwidth]{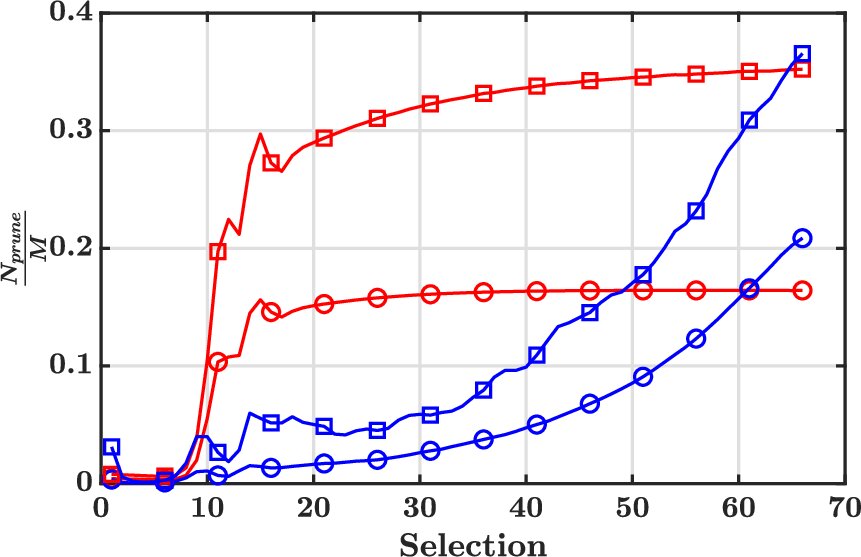}
\caption{Left: computational gain expressed in terms of Eq.~\ref{eq:compGain} when targeting the level $1356$ with fixed weight (\mythickbarredcircle{red}{white}) and time-dependent weight (\mythickbarredcircle{blue}{white}); when targeting the level $1737$ with fixed weight (\mythickbarredsquare{red}{white}) and time-dependent weight (\mythickbarredsquare{blue}{white}). The vertical line with \mythickbarredcircle{black}{white} denotes the probability corresponding to the level $1356$, and with \mythickbarredsquare{black}{white} denotes the probability corresponding to the level $1737$. Right: pruning ratio when targeting the level $1356$ with fixed weight (\mythickbarredcircle{red}{white}) and time-dependent weight (\mythickbarredcircle{blue}{white}); when targeting the level $1737$ with fixed weight (\mythickbarredsquare{red}{white}) and time-dependent weight (\mythickbarredsquare{blue}{white}). Plots correspond to the Lorenz 96 case with 32 degrees of freedom.}
\label{fig:perf_l96_32}
\end{center}
\end{figure}

The performances of the importance splitting algorithms with fixed and time-dependent weights are further investigated in Fig.~\ref{fig:perf_l96_32}, where the variance of the probability estimator is plotted in terms of the computational gain compared to a naive Monte-Carlo simulation. More precisely, the computational gain $G$ is defined as the number of brute force simulations required to achieve the same variance as the ISP, divided by the number of particles $M$ used for the ISP:

\begin{equation}
    G = \frac{p-p^2}{M \sigma_{ISP}^2},
    \label{eq:compGain}
\end{equation}

where $p$ is the probability of the rare event which is obtained with a Monte Carlo simulations, and $\sigma_{ISP}^2$ is the variance of the importance sampling estimator which is computed by repeating multiple times the ISP.

For both the fixed weight and time-dependent weighting factor, significant computational savings can be obtained, especially for low-probability events. The target probability corresponding to $Q(T_f)=1356$ and $Q(T_f)=1737$ are indicated as the vertical black lines. Interestingly, the importance splitting algorithm does not provide computational gains at these levels, which means that the final distribution of $Q$ is not centered around the target levels. Therefore, the choice of the weights according to the rationale of Eq.~\ref{eq:weight} does not guarantee computational efficiency at the level targeted. As expected, when the cloning is more aggressive, the computational gain increases more steeply as the probability decreases. This is observed for both the fixed and time-dependent weights. The main difference between fixed and time-dependent weighting is apparent when targeting higher values of the QoI. The fixed weight only provides computational gain for probabilities lower than $10^{-6}$ and is consistently outperformed by the time-dependent weight algorithm. This result can be explained by the fact that the fixed weight method discards many trajectories too quickly. The average pruning ratio taken across the $10^5$ realizations of the ISP for the fixed and time-dependent weighting factors are shown in Fig.~\ref{fig:perf_l96_32} for both target levels, and indicate a consistently larger pruning ratio for the fixed weight method. At lower target levels, while the pruning ratio is larger, the computational gains between the fixed and time-dependent weighting factors are almost the same. It is likely that even a small number of additional trajectories that are not pruned can have a disproportionate effect on the performance of the estimator. This further emphasizes the importance of the weighting method for this algorithm. Additional cases are run with 64 and 1024 degrees of freedom instead of 32 to evaluate the effect the dimensionality of the problem of the findings. The results (shown in \ref{app:dimension}) suggest that the above findings hold for higher number of dimensions as well.

\section{Self-similarity approach for estimating the path to a rare event}
\label{sec:estimateRarePath}

In the previous section, it was found that knowing the mean path to a rare event can significantly improve the statistics of the probability estimators, especially to estimate low probabilities. However, to obtain the rare mean path, the approach used in Sec.~\ref{sec:fluc} required to observe many rare events in the first place, which defeats the purpose of the procedure. Instead, it is preferable to estimate the path to a rare event using a different method. In this section, a brief review of the available methods is provided in order to convey the point that simpler procedures are needed. For this purpose, the self-similarity based approach is introduced here. The method is tested using both a priori calculations and a posteriori analysis.

As mentioned in Sec.~\ref{sec:rarePath}, it has been widely recognized that the knowledge of the average or most likely path leading to a rare event is advantageous in computing the probability of this event. In computational chemistry, transition paths are typically obtained through an optimal sequence of trials and errors, also referred to as the shooting method \cite{dellago2002transition}. Another approach well-suited for noise-driven rare events consists of finding the optimal time history of external forcing that can drive the system to a rare event. This optimum can be obtained by solving a minimization problem \cite{zhang2018rare,laurie2015computation} or by directly solving a partial differential equation (PDE) \cite{grafke2015instanton} for this instanton. However, such methods can create numerical issues for deterministic chaotic systems that evolve over long periods of time \cite{souza}. In this work, a simpler approach is developed, motivated by the observation that there exists self-similarity in the paths that lead to rare events for the system investigated.

\subsection{A priori analysis}
\label{sec:priori}

\begin{figure}[h]
\begin{center}
\includegraphics[width=0.45\textwidth]{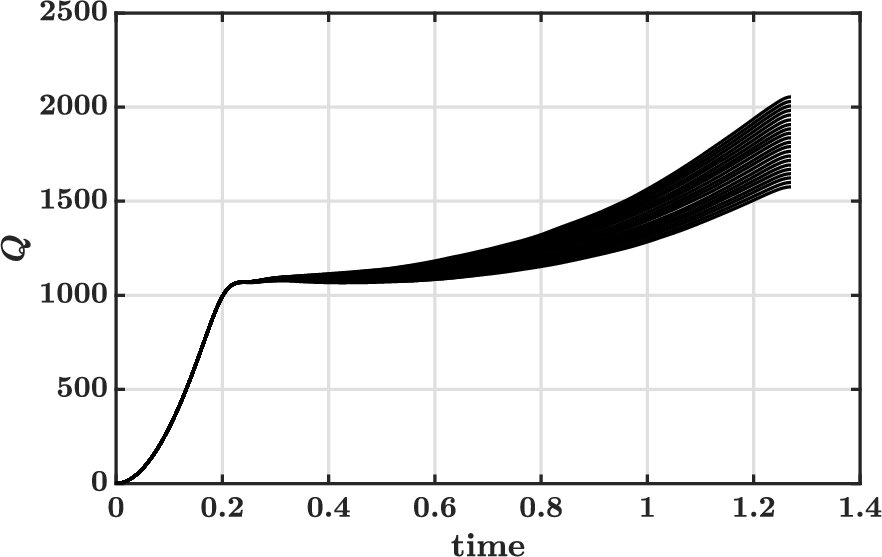}
\includegraphics[width=0.45\textwidth]{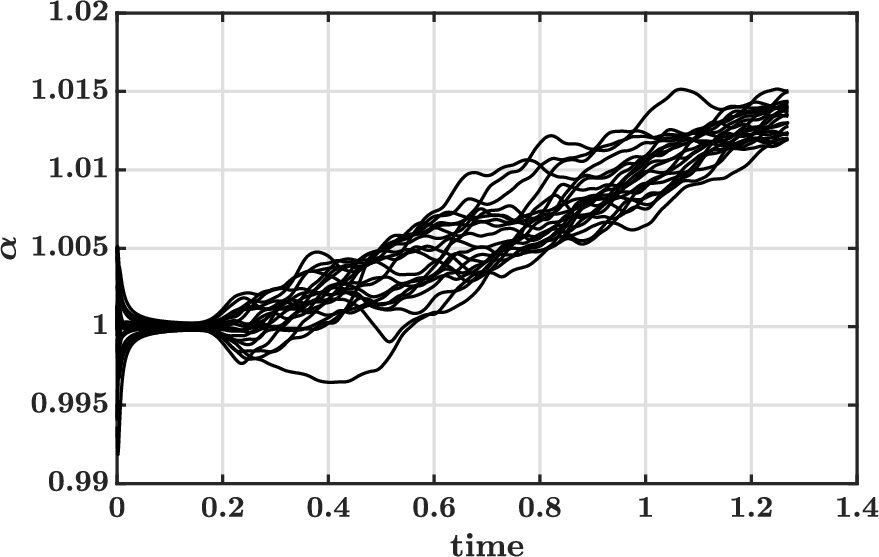}
\caption{Left: rare mean path exceeding thresholds ranging from $1500$ to $2000$ and separated by a stepsize of $25$. Right: self-similarity factor $\alpha$ computed from successive rare mean path (see Eq.~\ref{eq:selfsim}). Plots correspond to the Lorenz 96 case with 32 degrees of freedom.}
\label{fig:selfSim}
\end{center}
\end{figure}

Using the brute force Monte-Carlo simulation for the Lorenz 96 case with 32 degrees of freedom (discussed in Sec.~\ref{sec:fluc} with $8\times 10^{8}$ trajectories), the rare mean path to many target levels can be accurately estimated. Several thresholds levels are selected between $1500$ and $2000$ and are separated by a constant step of size $25$. The rare mean paths obtained are plotted in Fig.~\ref{fig:selfSim}.

It can first be observed that the path leading to each one of the thresholds are surprisingly similar to one another. In other words, the paths that leads to rare events share characteristics. This feature is termed \textit{self-similarity}, which provides an approach to estimating the path leading to a rare event without necessarily observing it. The simplest model that can be formulated would consist in assuming that the path leading to a higher threshold for the QoI is simply shifted up by a certain ratio. More formally, given a sequence of thresholds $a_1 < ... < a_m$ such that $\Delta a = a_{i+1}-a_i$ is constant and defining $R_i(t) = \langle Q(t)~|~Q(T_f) > a_i\rangle$,

\begin{equation}
    \forall i\in[1,m], \frac{R_{i+1}(t)}{R_i(t)}  = \alpha(t),
    \label{eq:selfsim}
\end{equation}
where $\alpha$ is a scalar independent of $i$. Here, $\alpha$ can be understood as a self-similarity factor that relates the rare mean paths separated by a step size of $\Delta a$ for the final value of the observable. In Fig.~\ref{fig:selfSim}, this model is a priori tested by plotting the value of $\alpha$. Indeed, the ratio $\alpha$ appears almost independent of the threshold considered which is in itself surprising given the simplicity of the model. It should be noted that many other models could be formulated and could involve the value of the QoI rescaled by its mean, or a more elaborate factor that depends on the response of the system to perturbation (Lyapunov exponents). Note that the same features are found with the Lorenz 96 system using 64 and 1024 degrees of freedom (\ref{app:dimension}).

\subsection{A posteriori analysis}
\label{sec:posteriori}

To test the effectiveness of the self-similarity approach, an a posteriori analysis is conducted for the Lorenz 96 case with 32 degrees of freedom. A Monte-Carlo calculation with $M=2500$ samples is conducted, where $M$ is also the number of particles used in the importance splitting algorithm. Note that it is not unreasonable to expect a Monte-Carlo calculation to be run with $M$ particles even for other purposes than of estimating the rare mean path. For example, a user of the ISP would typically run randomly sampled simulations in order to cross validate the output of the ISP, at least for the highest probability obtained with the ISP. Furthermore, the value of the weighting factor requires estimating the first and second moments of the QoI, for which randomly sampled simulations would have to be run. Additionally, given that the probability estimate from the ISP is subject to some variance, it is preferable to run the ISP multiple times. The computational cost of the rare event path estimation is therefore of the same order of magnitude as other sanity checks that the user needs to run in general. 

\begin{figure}[h]
\begin{center}
\includegraphics[width=0.30\textwidth]{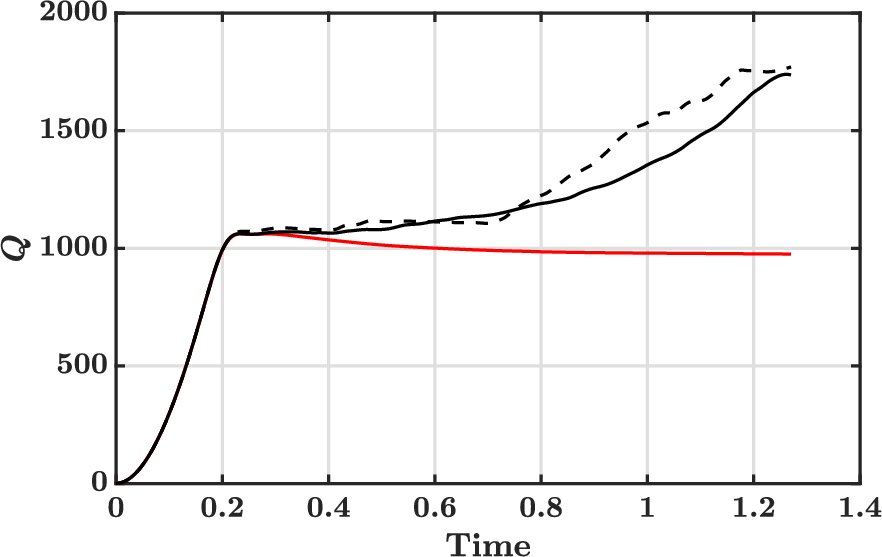}
\includegraphics[width=0.30\textwidth]{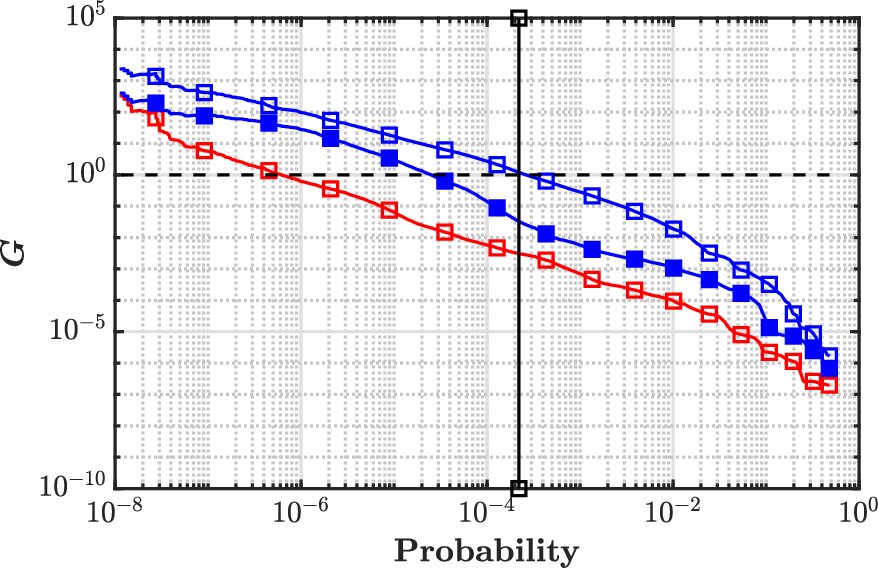}
\includegraphics[width=0.30\textwidth]{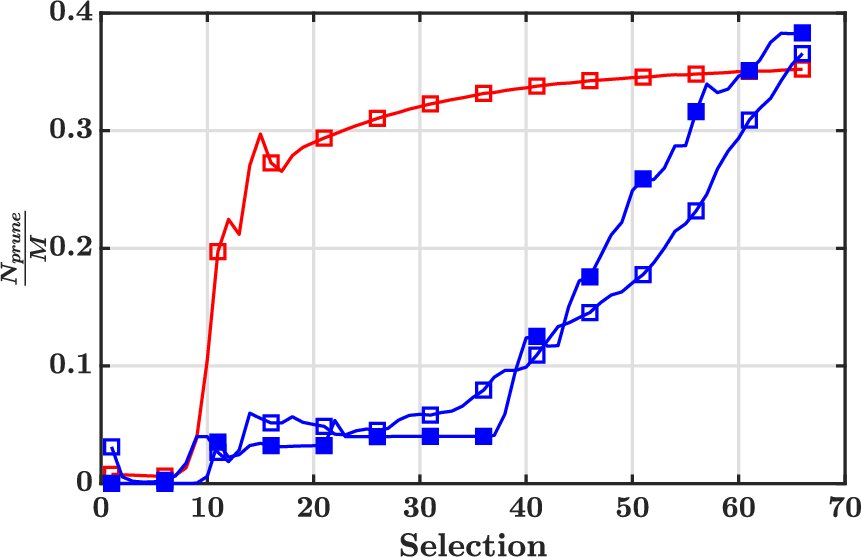}
\caption{Left: rare mean path $R_{1737}$ obtained from brute force computation (\mythickline{black}), from the self-similarity approximation (\mythickdashedline{black}) and ensemble average time history of the observable (\mythickline{red}). Middle: computational gain computed as Eq.~\ref{eq:compGain} when targeting the level $1737$ with the fixed weight (\mythickbarredsquare{red}{white}), the time-dependent weight obtained from the brute force calculation (\mythickbarredsquare{blue}{white}), the time-dependent weight obtained from the self-similarity approximation \mythickbarredsquare{blue}{blue}. Right: pruning ratio obtained when targeting the level $1737$ with fixed weight (\mythickbarredsquare{red}{white}) and time-dependent weight obtained from brute force calculation (\mythickbarredsquare{blue}{white}) and the time-dependent weight obtained from the self-similarity approximation (\mythickbarredsquare{blue}{blue}). Plots correspond to the Lorenz 96 case with 32 degrees of freedom.}
\label{fig:postperf}
\end{center}
\end{figure}

From the $2500$ brute force calculations, the mean trajectory to threshold ranging from $1225$ to $1325$ are directly computed (based on definition in Eq.~\ref{eq:defRareMeanPath}), in steps of $\Delta a = 25$, leading to at least $70$ trajectories per threshold. The path $R_{1737}(t)$ to the threshold $Q(T_f) = 1737$ is then simply estimated as 

\begin{equation}
    \forall t \in [0, T_f],~R_{a}(t) = \alpha_{ave}(t) ^{\frac{a-b}{\Delta a}} ~ R_{b}(t),
\end{equation}

where $R_a$ is the path to the target level (here, $a=1737$), $b$ is the largest level directly sampled from the $M$ random samples (here, $b=1325$), $\alpha_{ave} (t) = \langle \frac{R_{i+1}(t)}{R_{i}(t)} \rangle$.  From here, the  time-dependent weight $C(t)$ can be constructed according to Eq.~\ref{eq:weight}. However, the extrapolation procedure incurs numerical errors that can be  amplified in regions where $\sigma_Q^2(t)$ is close to 0. In practice, the absolute value of the weighting factor can be extremely large in these regions and lead to unreasonable pruning ratio. Instead, the weighting factor can be regularized by recognizing that the particle selection should be more and more aggressive close to $T_f$. In other terms, $|C(t)|$ should be an increasing function of time. Since the variance of the observable is expected to be large at $T_f$ (the perturbations have had sufficient time to amplify), the numerical error at the latest time is least pronounced. Therefore, $C(t)$ can be enforced to be an increasing function of time in the following manner: starting from the value of $C$ at the final time, step backwards from $t$ to the previous timestep $t-\Delta t$. If $C(t-\Delta t)>C(t)$ then $C(t-\Delta t)=C(t)$; otherwise, do nothing. Then keep stepping backwards (compare $C(t-\Delta t)$ and $C(t-2 \Delta t)$). This procedure can then lead to a step-like function for $C(t)$ which will be reflected in the pruning ratio (Fig.~\ref{fig:postperf}).

Figure~\ref{fig:postperf} shows the extrapolated path towards the desired threshold compared with path obtain from brute-force Monte Carlo computation. The extrapolation procedure does incur inaccuracies, but overall follows the correct trend. The ISP algorithm is then run with $M=2500$ particles, with the weights $C(t)$ obtained from the extrapolated mean path. The ISP is executed $10^5$ times in order to obtain good statistical convergence of the estimator. As can be seen in Fig.~\ref{fig:postperf}, the computational gain obtained with the a posteriori test is slightly lower than by using a brute force calculation to compute the rare mean path. This is because the rare mean path estimated using the self-similarity model is subject to inaccuracies, which lead pruning more observations than needed (See Fig.~\ref{fig:postperf} right). However, the a posteriori test still outperforms the method with the fixed weighting factor, across all probabilities. By examining the pruning ratio, it appears that fewer samples are pruned when using the self-similarity approach as compared to the fixed weight method. The pruning ratio exhibits a series of steps that are due to the correction of numerical errors in the extrapolation. The same procedure appears to hold for a 64-dimensional and 1024-dimensional system (\ref{app:dimension}).

\section{Applicability to other systems}
\label{sec:applicability}

In Sec.~\ref{sec:rarePath} it was shown that the knowledge of the rare mean path could improve the convergence properties of the ISP algorithm. In Sec.~\ref{sec:estimateRarePath}, it was found that a simple approach could be used to estimate the rare mean path from less rare mean paths, and was demonstrated in the case of the Lorenz 96 case. In this section, it is examined whether the observations made in Sec.~\ref{sec:estimateRarePath} are applicable for problems more representative of turbulent flows, which is the motivation behind this work. An a priori analysis similar to the one conducted in Sec.~\ref{sec:estimateRarePath} is conducted for two different dynamical systems. 

\subsection{Kuramoto-Sivashinsky Equation (KSE)}

\begin{figure}[h]
\begin{center}
\includegraphics[width=0.45\textwidth]{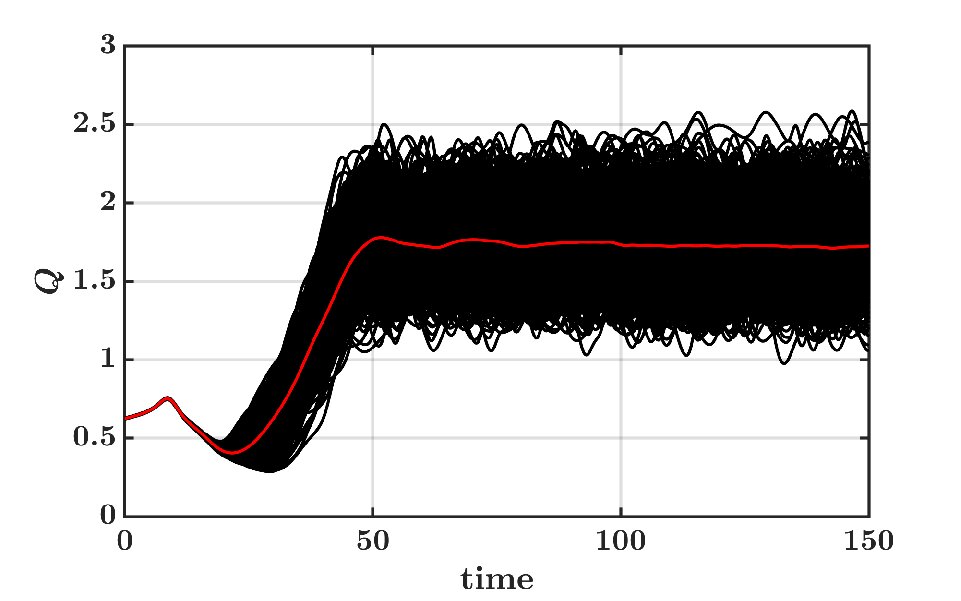}
\includegraphics[width=0.45\textwidth]{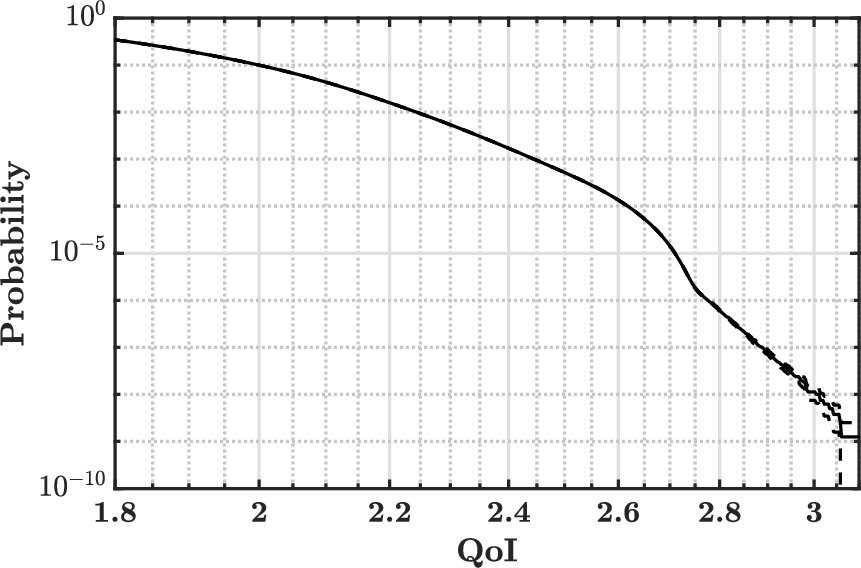}
\caption{Left: instantaneous values of the observable $Q(t)$ plotted over time (\mythickline{black}). Ensemble average of $Q$ (\mythickline{red}). Right: complementary of the cumulative density function (CDF) of the QoI. Probabilities are obtained with brute force calculation (\mythickline{black}) plotted along with the theoretical Monte-Carlo uncertainty (\mythickdashedline{black}). Plots correspond to the KSE case.}
\label{fig:kseQoI}
\end{center}
\end{figure}

The 1D Kuramoto-Sivashinsky \cite{kuramoto,sivashinsky} equation is often used as a surrogate for the spatiotemporal chaos seen in turbulent flows. In this work, the formulation with with unit viscosity coefficient is used:
\begin{equation}
    \forall t \in [0, T_f], \frac{\partial u}{\partial t} + \nabla^4 u + \nabla^2 u + \nabla u^2 = 0,
\end{equation}
where $u$ is defined on the domain $[0, 32 \pi]$ and $u(t=0) = cos(x/16) \cdot (1+sin(x/16))$, and $T_f = 150$ time units. The equations are integrated in time using the ETDRK4 scheme \cite{etdrk4} with a fixed timestep of $0.25$ time units. The equations are solved in Fourier space using $N=128$ modes. Here, $\bbxi$ is defined as the discrete version of $u$ in physical space, i.e.,  $\bbxi \in \mathbb{R}^{128}$ and the function $q$ that defines the QoI is defined as 
\begin{equation}
    Q(t) = \frac{1}{N} \sum_{i=1}^N \xi_i^2.
\end{equation}
Figure~\ref{fig:kseQoI} shows the evolution of the QoI for different initial conditions, and exhibits variations representative of a chaotic system. The event probability is then defined as $P(Q(T_f)>a)$. The probabilities corresponding to different thresholds are shown in Fig.~\ref{fig:kseQoI} (right) and were obtained by running $8 \times 10^8$ trajectories.

\begin{figure}[h]
\begin{center}
\includegraphics[width=0.45\textwidth]{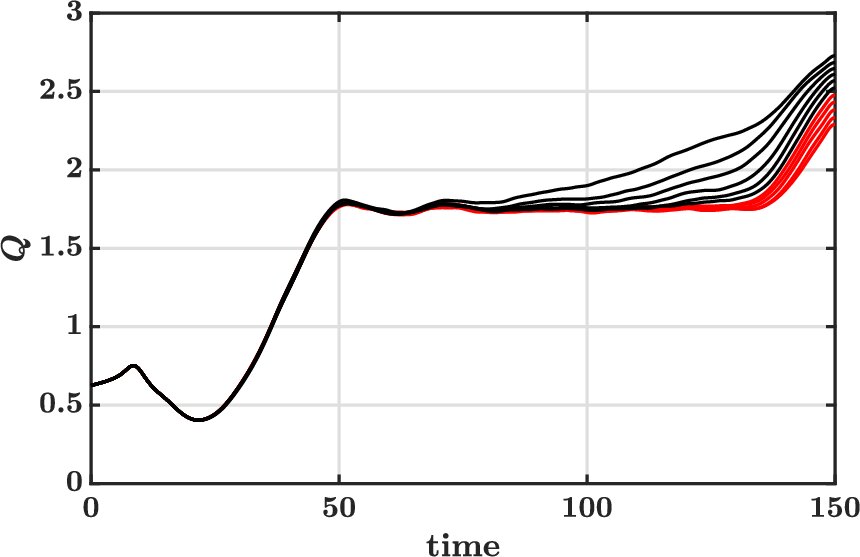}
\includegraphics[width=0.45\textwidth]{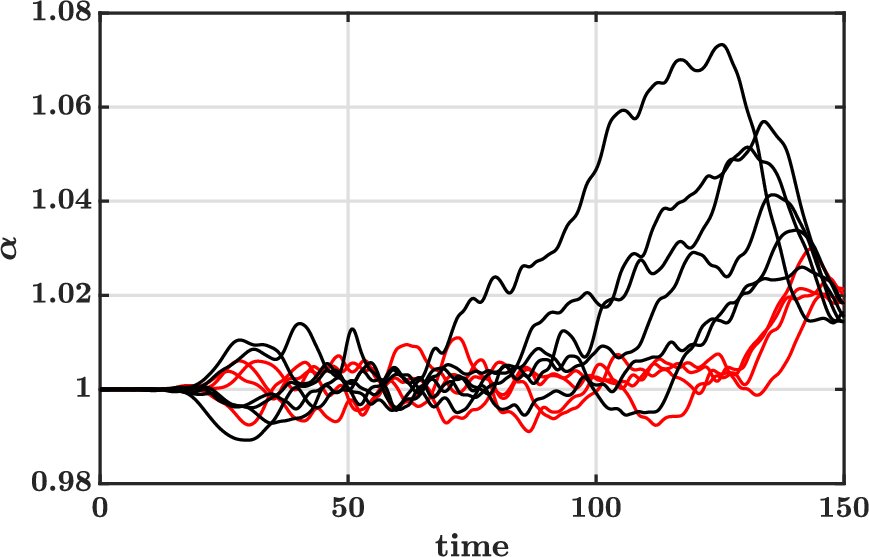}
\caption{Left: rare mean path exceeding thresholds ranging from $2.2$ to $2.7$ and separated by a stepsize of $0.05$. Right: self-similarity factor $\alpha$ computed from successive rare mean path (see Eq.~\ref{eq:selfsim}). Plots correspond to the KSE case.}
\label{fig:aprioriKS}
\end{center}
\end{figure}

The rare mean path for thresholds ranging from $2.2$ to $2.7$ (thresholds for which at least 1000 trajectories can be obtained) and separated by a constant step of $0.05$ are shown in Fig.~\ref{fig:aprioriKS} (left). It can be observed that the rare mean path follow the same functional at first (red lines), which suggest that the self-similarity approach holds initially. However, the rarest mean paths gradually deviate from the initial functional and would not be reasonably estimated using the self-similarity approach. In Fig.~\ref{fig:aprioriKS} (right), the self-similarity factor $\alpha(t)$ is plotted. The self-similarity approximation is reasonably accurate at early to intermediate times, but shows more variability at later times. Overall, these observations suggest that the self-similarity approach could be useful for short term predictions. Nevertheless, the rare mean paths do not significantly differ from one another, and it is still reasonable to expect the less rare mean paths to be informative of the rarer mean path. 

\subsection{High-altitude relight in aircraft engines}

In aircraft engines operating at high altitude, there is a finite possibility of the flame blowing out, which will lead to loss of propulsion. For this reason, the ability to relight the combustor within short times is crucial for safety (and necessary to obtain certification). During the relight procedure, a spark source is repeatedly used to inject high enthalpy gases into a combustor that is fueled but is at low temperature. While some of these sparks can lead to ignition, propagation of the flame kernel, and eventual stabilization of the gas turbine, the chaotic flow inside the combustor introduces a source of uncertainty. The goal then is to estimate the probability of ignition given a set of operating conditions. 

\begin{figure}[h]
\begin{center}
\includegraphics[width=0.9\textwidth]{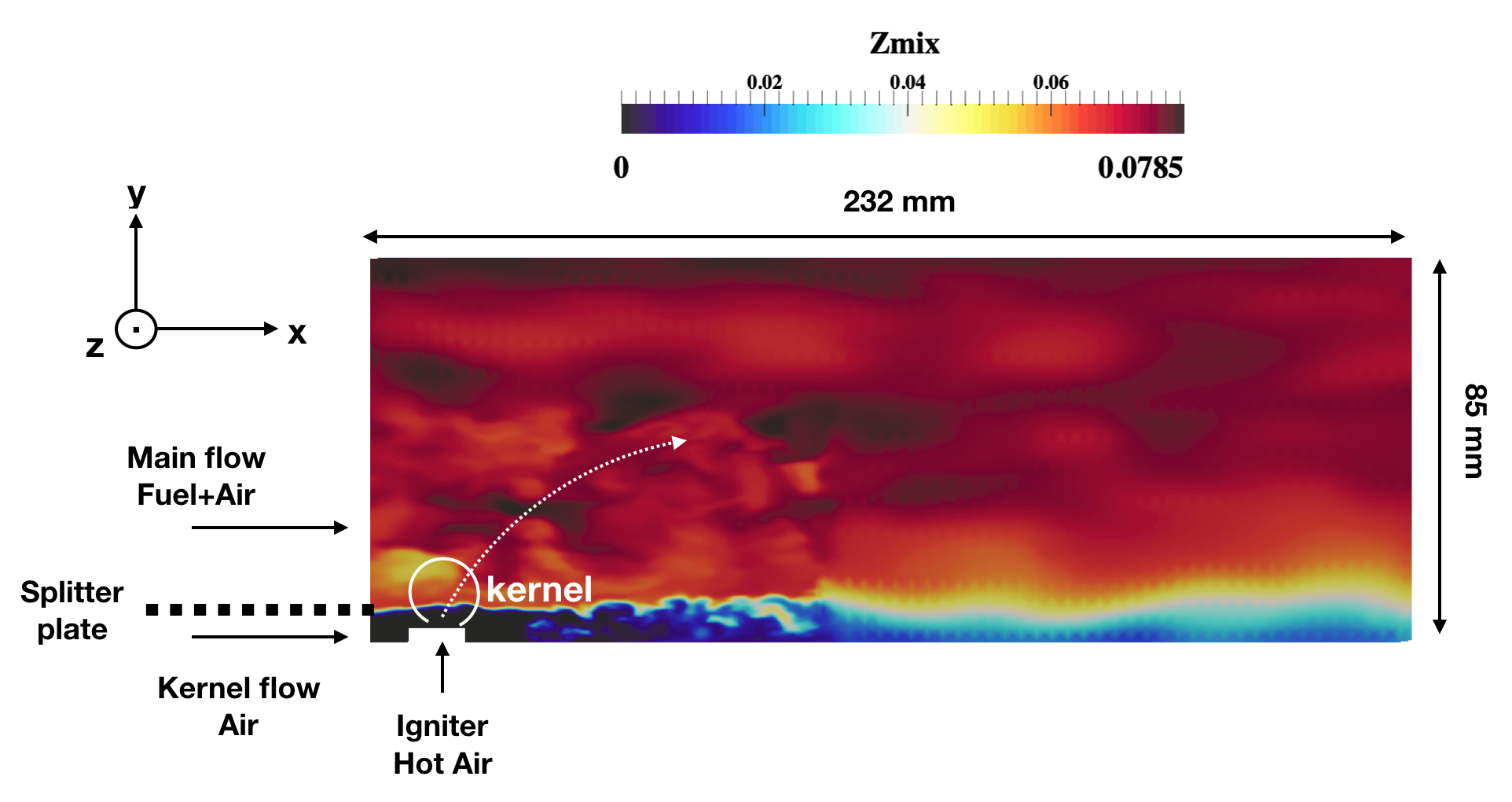}
\caption{Schematic of the ignition configuration simulated. The contour of an initial mixture fraction field at the spanwise mid-plane of the configuration is shown. At the initial time, a hot kernel is ejected from the bottom and mixes with surrounding fuel after it emerges. The trajectory of a spark kernel is sketched by the dashed white line.}
\label{fig:schematicIgnition}
\end{center}
\end{figure}

A canonical flow configuration that replicates this ignition process was experimentally studied by \cite{gatechignition} and the a corresponding simulation is shown in Fig.~\ref{fig:schematicIgnition}, based on the modeling study of \cite{tang2019comprehensive}.  Here, the spark is injected from the base of the flow (marked igniter in the figure). The kernel then traverses into the region that contains the fuel-air mixture and convecting in a turbulent flow. Subsequent mixing and chemical reactions lead to either a successful or failed ignition event. The forward model is deterministic in nature, similar to other examples studied in the current work. The main sources of uncertainty are the discharge efficiency of the spark igniter and the initial turbulence state in the system. It was determined that the turbulence effect is dominant over igniter efficiency for the range of conditions considered. To quantify ignition, the volume of the kernel at a particular time is used to mark the outcome of the spark injection. More precisely, the observable $Q$ is the volume of burnt products, and the rare event is defined in terms of the final volume of the ignition kernel, such that $P(Q(T_f)<a)$ needs to be estimated. It is emphasized that conversely to the KSE and L96 examples, lower thresholds correspond to lower probability. Hence, $R_{i+1}(t)< R_i$ and $\alpha <1$.

Details of the simulation procedure and operating conditions are provided in \cite{malikmcs11}. Briefly, each simulation of a sparking event is conducted using the large eddy simulation (LES) procedure, with an initial flow field that is sampled from a well-developed homogeneous flow. A total of 541 calculations are conducted among which 235 led to ignition success. Note that the number of simulations conducted is much lower than for the other cases since the system contains approximately $10^7$ degrees of freedom.

\begin{figure}[h]
\begin{center}
\includegraphics[width=0.45\textwidth]{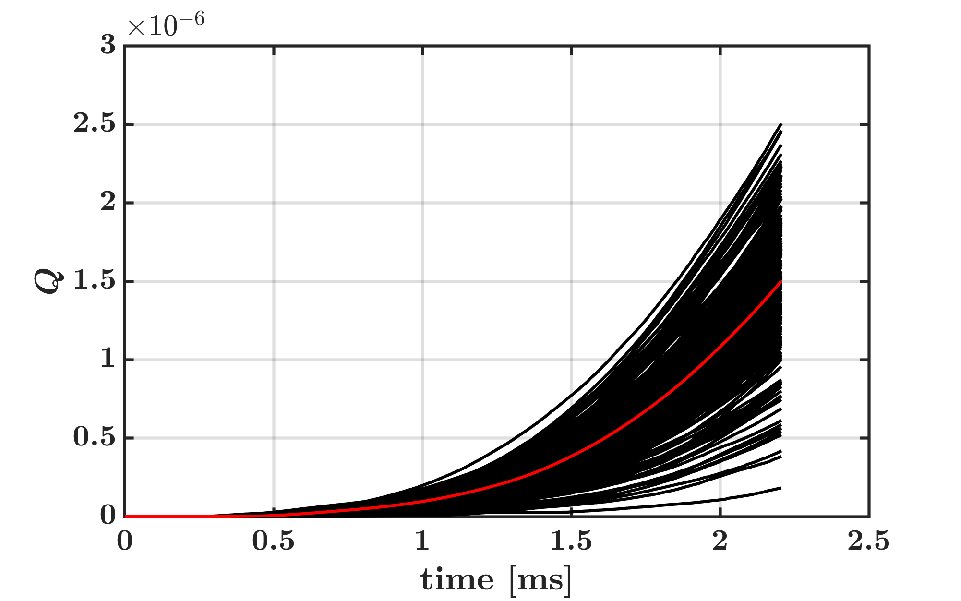}
\includegraphics[width=0.45\textwidth]{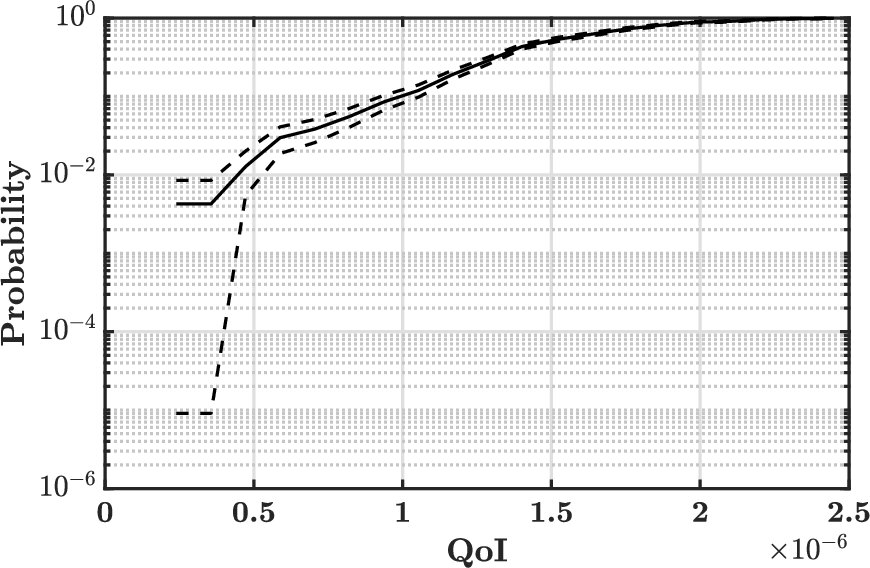}
\caption{Left: instantaneous values of the observable $Q(t)$ plotted over time (\mythickline{black}). Ensemble average of $Q$ (\mythickline{red}). Right: cumulative density function (CDF) of the QoI. Probabilities are obtained with brute force calculations (\mythickline{black}) plotted along with the theoretical Monte-Carlo uncertainty (\mythickdashedline{black}). Plots correspond to the high-altitude relight case.}
\label{fig:ignitionQoI}
\end{center}
\end{figure}

The results of the calculations are shown in Fig.~\ref{fig:ignitionQoI}. The volume of the kernel appears to grow exponentially, driven by the Arrhenius-type reaction chemistry that leads to ignition. As can be seen, all the calculations follow an exponential-like path. However, some ignition events result in weaker flames than others and are therefore more susceptible to be extinguished at later times. The CDF of the volume of the ignition kernel is shown in Fig.~\ref{fig:ignitionQoI} (right) and can only be reasonably estimated for probabilities larger than $10^{-2}$, due to the small sample set used.

\begin{figure}[h]
\begin{center}
\includegraphics[width=0.45\textwidth]{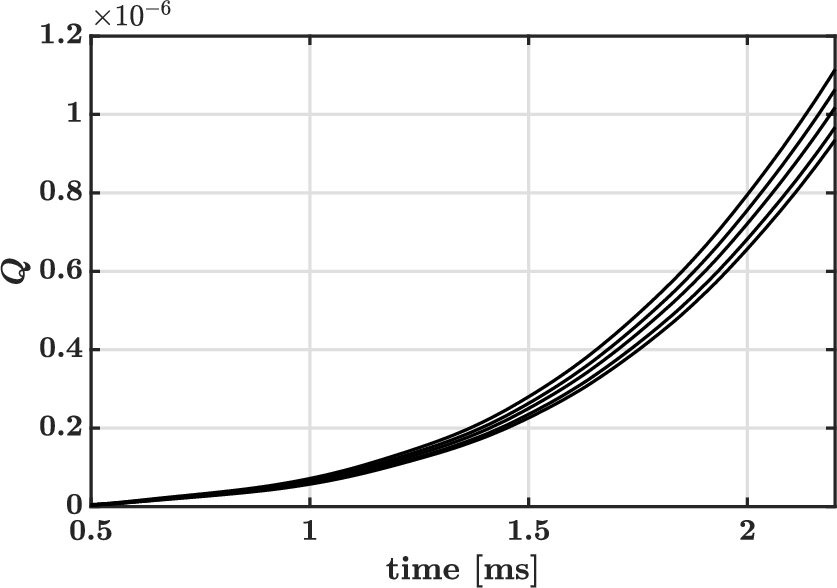}
\includegraphics[width=0.45\textwidth]{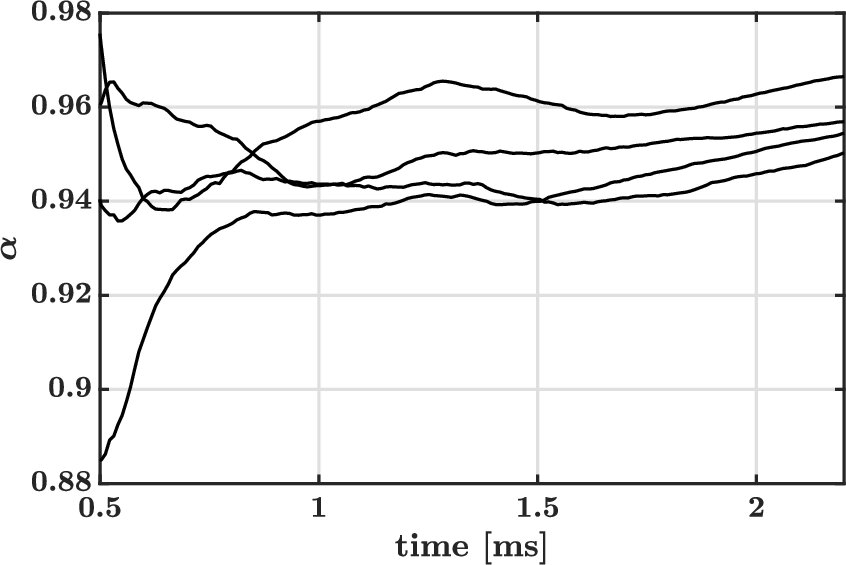}
\caption{Left: rare mean path exceeding thresholds ranging from $1.4$~cm$^3$ to $1.2$~cm$^3$ and separated by a stepsize of $0.05$~cm$^3$. Right: self-similarity factor $\alpha$ computed from successive rare mean path (see Eq.~\ref{eq:selfsim}). Plots correspond to the high-altitude relight case.}
\label{fig:aprioriIgnition}
\end{center}
\end{figure}

In Fig.~\ref{fig:aprioriIgnition}, the self-similarity approach is tested using a sequence of thresholds ranging from $a=1.4$~cm$^3$ to $a=1.2$~cm$^3$ with steps of $0.05$~cm$^3$. Note that the plots are only shown after $0.5$~ms as the kernel volumes are too small before that time and produce numerical errors when ratios are computed. The rare mean path follows a self-similar structure, which indicates that extreme events can be estimated from the less-rare mean paths. Figure~\ref{fig:aprioriIgnition} tests the self-similarity approach outlined in Sec.~\ref{sec:estimateRarePath}. At early times, the similarity factor shows large variations, but quickly reaches a nearly constant value in the range of $0.94-0.96$. The initial variability can be safely removed from the pruning process if the weighting factor is enforced to be a monotonically increasing function of time.  At later times, the values of $\alpha$ are close for the different threshold values, indicating that the self-similarity assumption holds for this system as well. 

\section{Conclusions}
\label{sec:conclusions}

Based on prior work of Wouters et al.~\cite{wouters_bouchet}, the computational advantage obtained by estimating the mean path to a rare event was demonstrated. It was shown that such a path estimation consistently reduces the variance of the probability estimator, especially when the goal is to compute very low probabilities.

Analysis of paths for events at increasingly rare QoIs shows that there exists a self-similarity in these paths that could be exploited to reduce computational cost. In particular, the path followed by a less rare event was found to have a form that is similar to that of a more rare event. An algorithm that exploits this characteristic was constructed and tested a priori and a posteriori for the Lorenz 96 case. It was shown to successfully outperform the importance splitting algorithm that uses a constant weighting factor. 

The existence of this self-similarity structure was studied in more complex cases. Two cases, namely the KSE and a high-altitude relight problem, were considered. In the KSE-based system, it was shown that self-similarity holds for a range of rare event probabilities, but can lead to errors when used for events with very low probability. The high-altitude relight problem showed that the self-similarity factor varies initially, but converges to a nearly constant value with time. Both these results demonstrate that the self-similarity approach is valid for deterministic problems, and can be extended to more complex systems. 

For future work, the self-similarity validity needs to be tested a posteriori for practical problems, such as the high-altitude relight configuration. This step is not necessarily straightforward as cloning a deterministic simulation requires that the individual trajectory is perturbed without affecting the rare event probability. In the case of a Lorenz 96 problem, a simple random cloning strategy could be used. For turbulent flows, where the spatial coherence of perturbations matter, a more detailed perturbation method is required. Another path for improvement would be to revisit the self-similarity model formulated in Sec.~\ref{sec:estimateRarePath}. At the moment, a rudimentary model that exploits the self-similarity property for the rare mean path was formulated. More elaborate models could be formulated to be more suited for different systems and to be more resilient to numerical errors. While it was shown that the present approach could be implemented with low computational overhead, it could be advantageous to leverage multi-fidelity approaches to learn the rare mean path, in the same vein as Ref.~\cite{peherstorfer2016multifidelity}. Finally, throughout this work, no distinction is made between the observable and the QoI. It could be advantageous to derive a method allowing to pick an observable for which self-similarity properties are more pronounced.

\section{Acknowledgements}
This work was financially supported by an AFOSR research grant (FA9550-15-1-0378) with Chiping Li as program manager. The authors thank NASA HECC for generous allocation of computing time on NASA Pleiades machine. The contribution of Yihao Tang in running the ignition simulations is also gratefully acknowledged.

\appendix

\section{Validation of algorithm for higher-dimensional problems}
\label{app:dimension}

\begin{figure}[h]
\begin{center}
\includegraphics[width=0.45\textwidth]{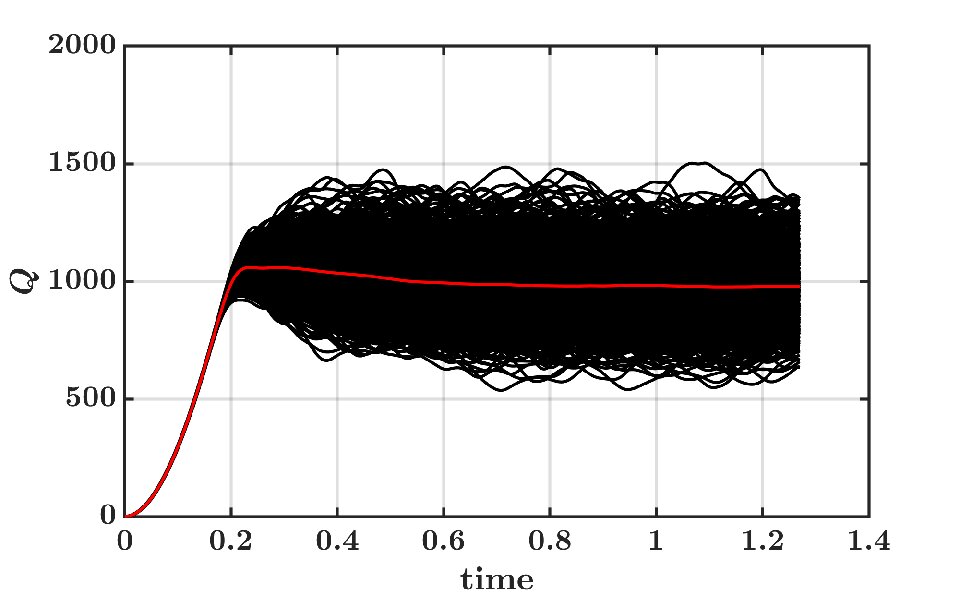}
\includegraphics[width=0.45\textwidth]{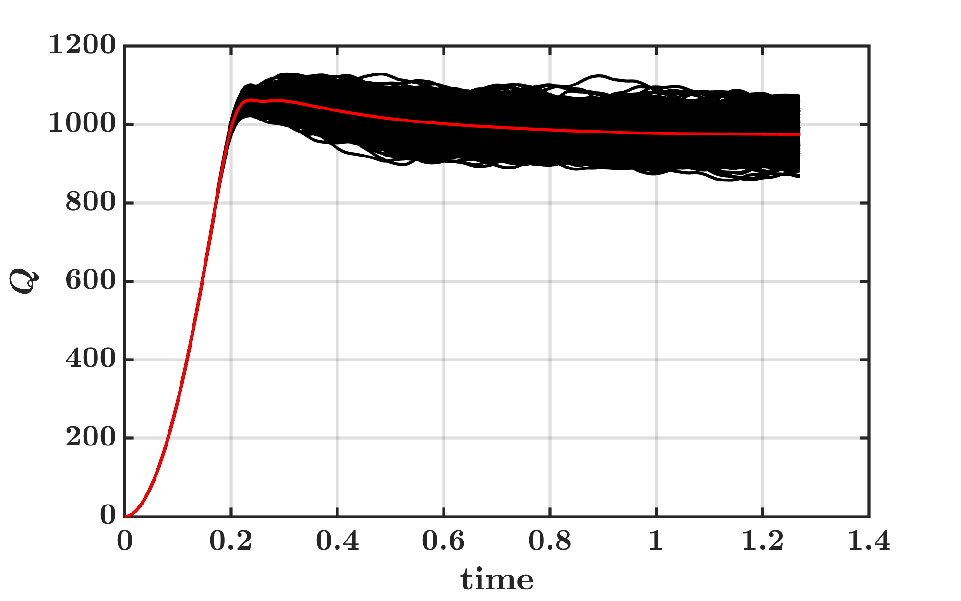}
\caption{Instantaneous values of the observable $Q(t)$ plotted over time (\mythickline{black}). Ensemble average of $Q$ (\mythickline{red}). Left: Lorenz 96 case with 64 degrees of freedom. Right: Lorenz 96 case with 1024 degrees of freedom.}
\label{fig:illL9664}
\end{center}
\end{figure}

In this section, the Lorenz 96 system is investigated in 64 dimensions and 1024 dimensions to determine whether the features found at 32 dimensions hold with increase in dimensionality. The QoI is defined in the same manner as the levels chosen for the 32-dimensional case (Eq.~\ref{eq:qoi}). The numerical integration and the timestep size are the same as for the 32-dimensional case. Figure~\ref{fig:illL9664} shows the instantaneous evolution of the QoI for 1000 realizations, along with the ensemble average QoI, for the 64 and 1024-dimensional case. %Note that the variance of the QoI is smaller than in the 32 dimension case. Therefore when comparing the performances of the ISP for different number of dimensions, the target level will be adjusted so that it is located the same number of dimensions away from the mean, for all the n For comparison with the 32 dimensional case, the thresholds are chosen the same number of dimension away from  and will, therefore, require choosing threshold levels different than in the 32-dimensional case \alert{This does not make sense given the statement above - clarify}. 

\begin{figure}[h]
\begin{center}
\includegraphics[width=0.30\textwidth]{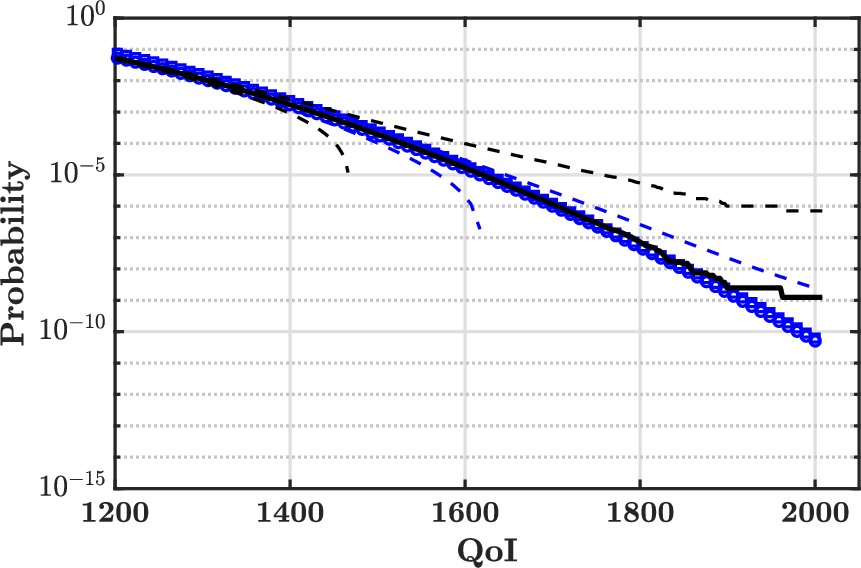}
\includegraphics[width=0.30\textwidth]{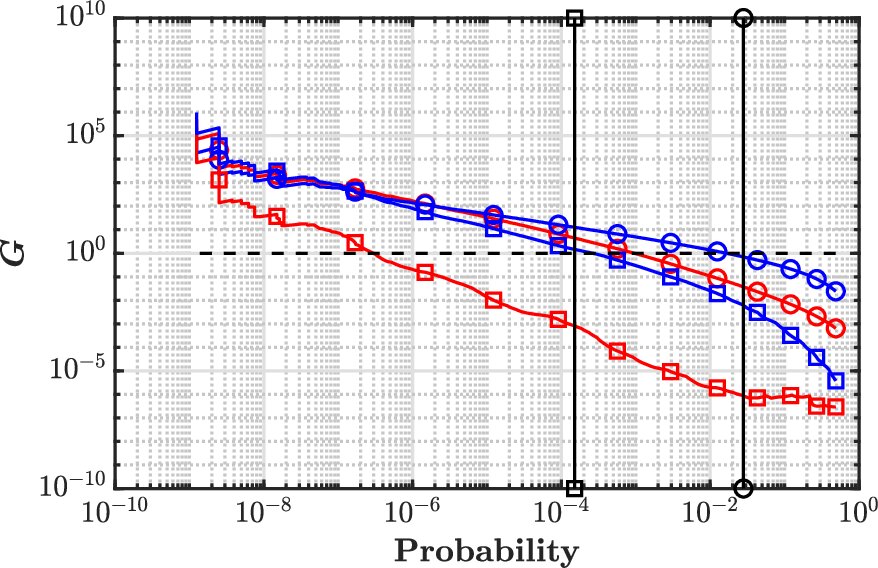}
\includegraphics[width=0.30\textwidth]{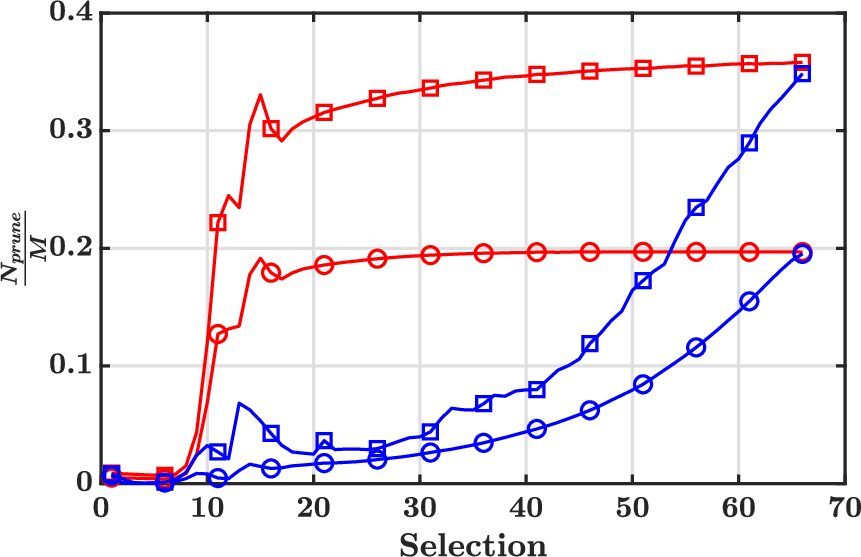}
\caption{Left: complementary of the cumulative density function (CDF) of the QoI. Probabilities are obtained with brute force calculation (\mythickline{black}), the ISP method with time-dependent weight based on $R_{1244}$ (\mythickcircle{blue}{white}), and $R_{1512}$ (\mythicksquare{blue}{white}) and $M=2500$ particles. Uncertainty of the estimator computed with the ISP algorithm using $R_{1244}$ for the time-dependent weight is shown (\mythickdashedline{blue}) along with the theoretical Monte-Carlo uncertainty that would be obtained with $M=2500$ realizations (\mythickdashedline{black}). Middle: computational gain computed as Eq.~\ref{eq:compGain} when targeting the level $1512$ with the fixed weight (\mythickbarredsquare{red}{white}), the time-dependent weight obtained from the brute force calculation (\mythickbarredsquare{blue}{white}), the time-dependent weight obtained from the self-similarity approximation \mythickbarredsquare{blue}{blue}. Right: pruning ratio obtained when targeting the level $1512$ with fixed weight (\mythickbarredsquare{red}{white}) and time-dependent weight obtained from brute force calculation (\mythickbarredsquare{blue}{white}) and the time-dependent weight obtained from the self-similarity approximation (\mythickbarredsquare{blue}{blue}). Plots correspond to the Lorenz 96 problem with 64 degrees of freedom.}
\label{fig:perf_l96_64}
\end{center}
\end{figure}

\begin{figure}[h]
\begin{center}
\includegraphics[width=0.30\textwidth]{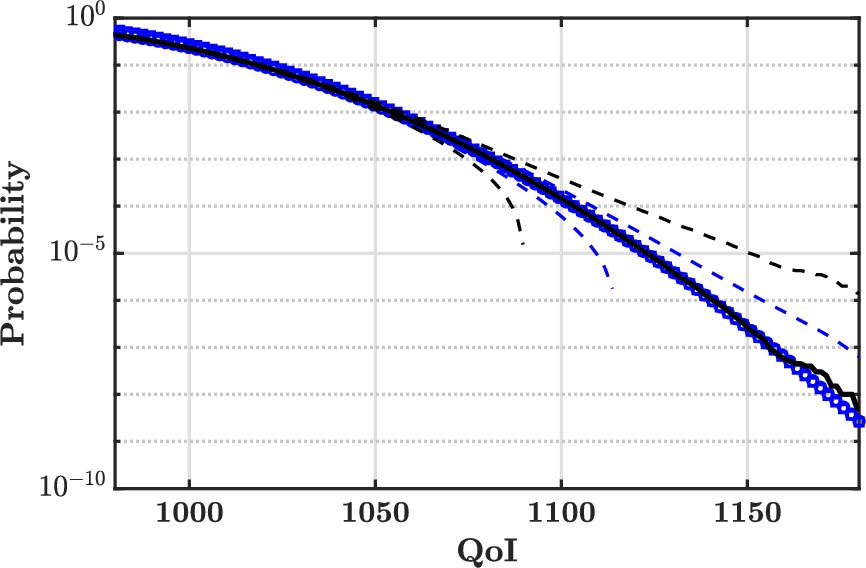}
\includegraphics[width=0.30\textwidth]{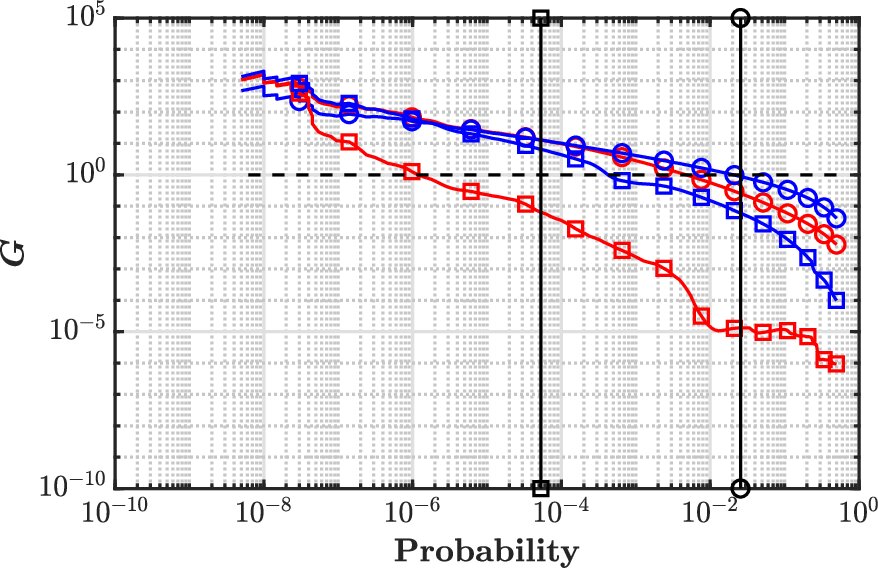}
\includegraphics[width=0.30\textwidth]{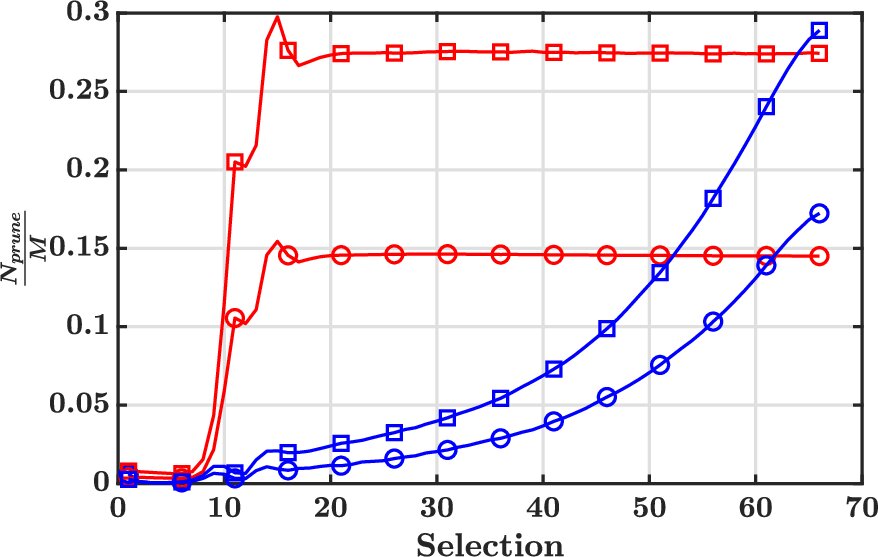}
\caption{Left: complementary of the cumulative density function (CDF) of the QoI. Probabilities are obtained with brute force calculation (\mythickline{black}), the ISP method with time-dependent weight based on $R_{1042}$ (\mythickcircle{blue}{white}), and $R_{1109}$ (\mythicksquare{blue}{white}) and $M=2500$ particles. Uncertainty of the estimator computed with the ISP algorithm using $R_{1042}$ for the time-dependent weight is shown (\mythickdashedline{blue}) along with the theoretical Monte-Carlo uncertainty that would be obtained with $M=2500$ realizations (\mythickdashedline{black}). Middle: computational gain computed as Eq.~\ref{eq:compGain} when targeting the level $1109$ with the fixed weight (\mythickbarredsquare{red}{white}), the time-dependent weight obtained from the brute force calculation (\mythickbarredsquare{blue}{white}), the time-dependent weight obtained from the self-similarity approximation \mythickbarredsquare{blue}{blue}. Right: pruning ratio obtained when targeting the level $1109$ with fixed weight (\mythickbarredsquare{red}{white}) and time-dependent weight obtained from brute force calculation (\mythickbarredsquare{blue}{white}) and the time-dependent weight obtained from the self-similarity approximation (\mythickbarredsquare{blue}{blue}). Plots correspond to the Lorenz 96 problem with 1024 degrees of freedom.}
\label{fig:perf_l96_1024}
\end{center}
\end{figure}

The target levels chosen for the 64 and 1024-dimensional cases are located two and four standard deviations away from the mean (similar to the choices made for the 32 dimensional case). The target levels chosen are $1244$ and $1512$ for the 64-dimensional case; $1042$ and $1109$ for the 1024-dimensional case. The number of particles is $M=2500$. All the ISP simulations are run $10^5$ times in order to ensure convergence of the statistics of the probability estimator. In Fig.~\ref{fig:perf_l96_64} (for the 64-dimensional case) and Fig.~\ref{fig:perf_l96_1024} (for the 1024-dimensional case), it can be seen that the estimators obtained are still unbiased (left plots) and provide a consistent improvement compared to the fixed weight methods (middle plots). This improvement can again be linked to the smaller pruning ratio induced by the time-dependent weighting factor.

\begin{figure}[h]
\begin{center}
\includegraphics[width=0.45\textwidth]{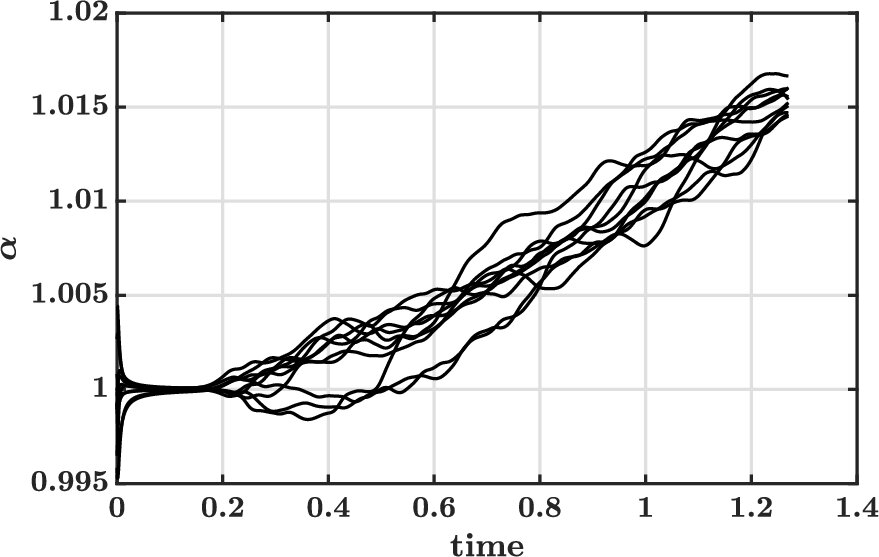}
\includegraphics[width=0.45\textwidth]{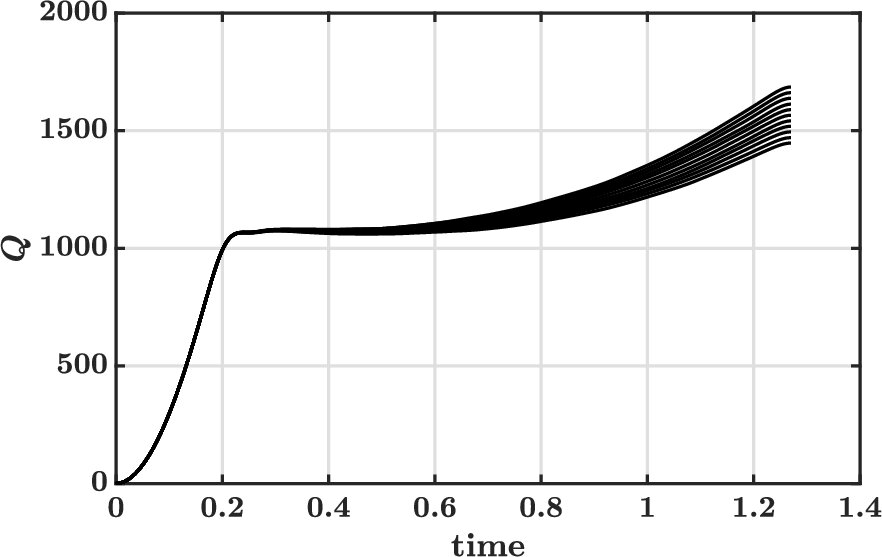}
\caption{Left: rare mean path exceeding thresholds ranging from $1400$ to $1650$ and separated by a stepsize of $25$. Right: self-similarity factor $\alpha$ computed from successive rare mean path (see Eq.~\ref{eq:selfsim}). Plots correspond to the Lorenz 96 case with 64 degrees of freedom.}
\label{fig:priori64}
\end{center}
\end{figure}

\begin{figure}[h]
\begin{center}
\includegraphics[width=0.45\textwidth]{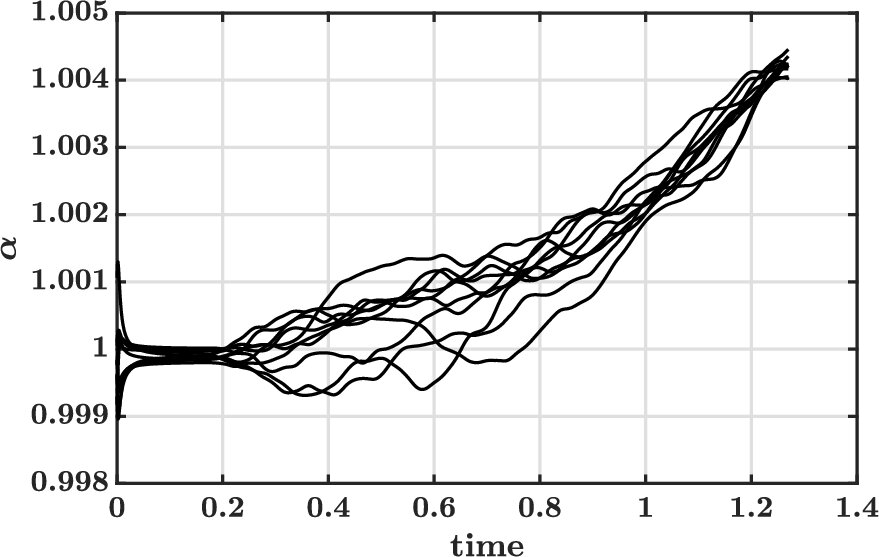}
\includegraphics[width=0.45\textwidth]{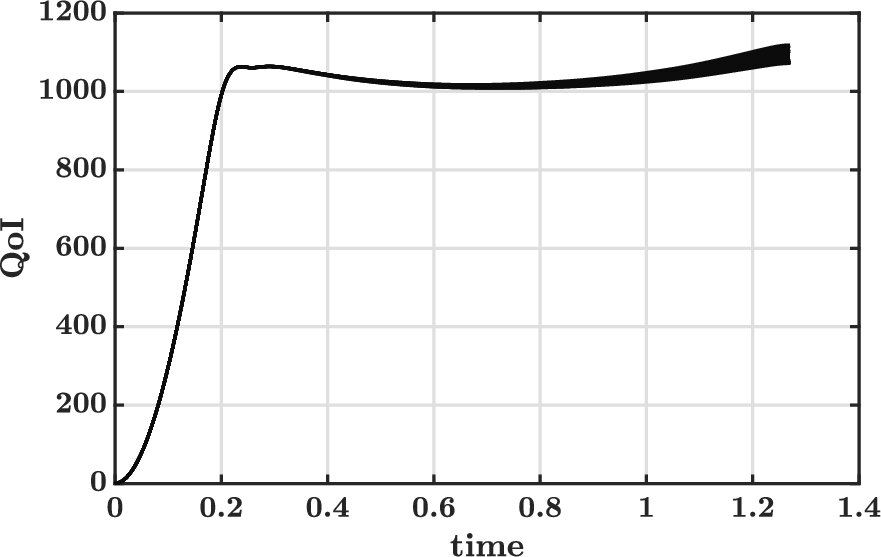}
\caption{Left: rare mean path exceeding thresholds ranging from $1060$ to $1110$ and separated by a stepsize of $5$. Right: self-similarity factor $\alpha$ computed from successive rare mean path (see Eq.~\ref{eq:selfsim}). Plots correspond to the Lorenz 96 case with 1024 degrees of freedom.}
\label{fig:priori1024}
\end{center}
\end{figure}

The self-similarity approach is tested a priori for the 64-dimensional case by sampling $8 \times 10^8$ realizations and computing the rare mean path exceeding levels ranging from $1400$ to $1650$ and separated by a constant step size of $25$. For the 1024-dimensional case, $2 \times 10^8$ realizations are run and rare mean paths exceeding levels ranging from $1060$ to $1110$ and separated by a constant step size of $5$ are shown. The results of the a priori analysis are shown in Fig.~\ref{fig:priori64} (for the 64-dimensional case) and Fig.~\ref{fig:priori1024} (for the 1024-dimensional case). Similar to the results obtained in Sec.~\ref{sec:priori}, there is a self-similar structure to these paths. The self-similarity model proposed in Sec.~\ref{sec:priori} is tested, and shows an equally valid level of agreement.

\begin{figure}[h]
\begin{center}
\includegraphics[width=0.30\textwidth]{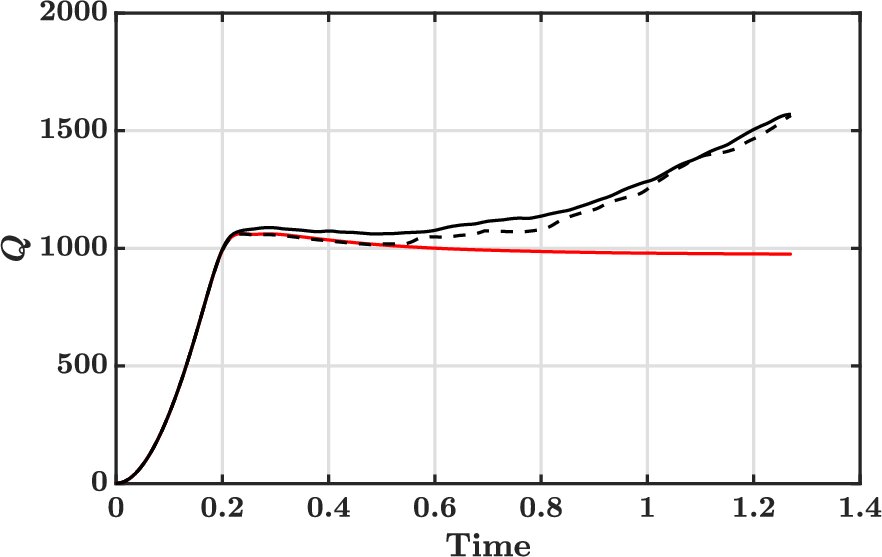}
\includegraphics[width=0.30\textwidth]{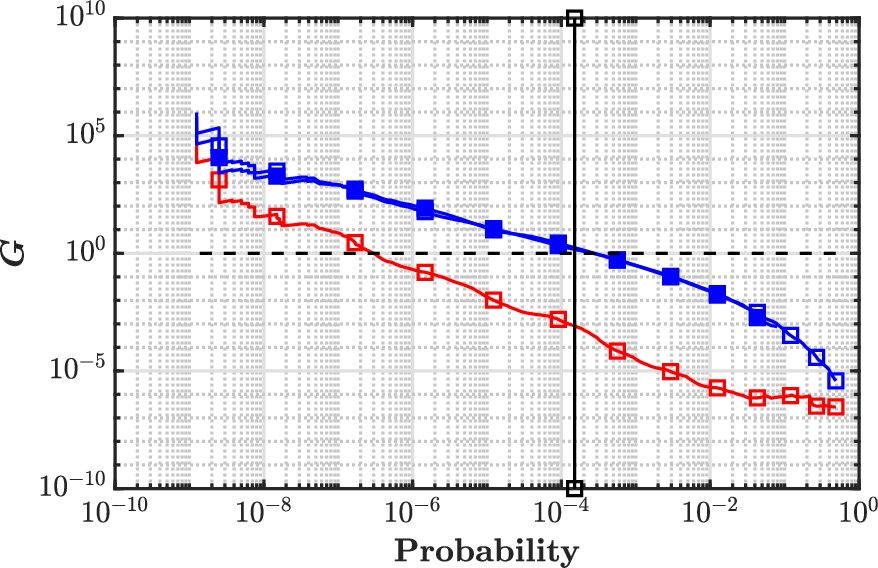}
\includegraphics[width=0.30\textwidth]{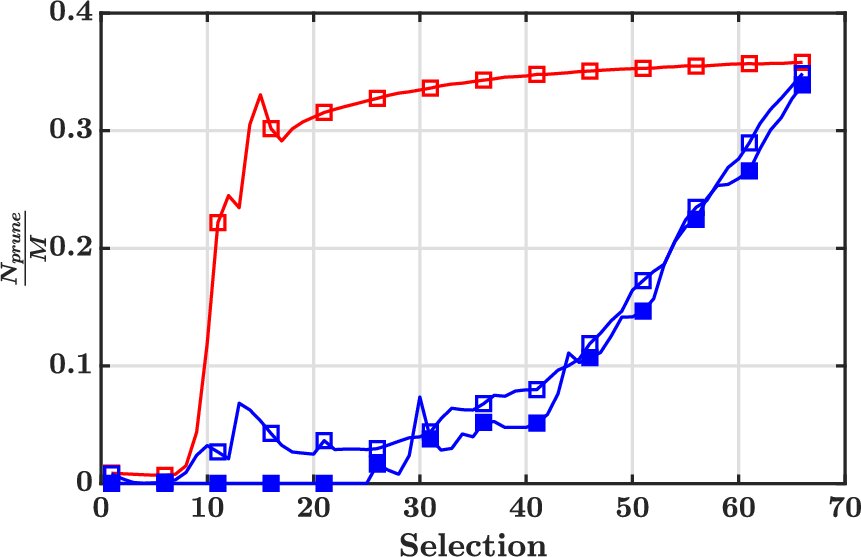}
\caption{Left: rare mean path $R_{1512}$ obtained from brute force computation (\mythickline{black}), from the self-similarity approximation (\mythickdashedline{black}) and ensemble average time history of the observable (\mythickline{red}). Middle: computational gain computed as Eq.~\ref{eq:compGain} when targeting the level $1512$ with the fixed weight (\mythickbarredsquare{red}{white}), the time-dependent weight obtained from the brute force calculation (\mythickbarredsquare{blue}{white}), the time-dependent weight obtained from the self-similarity approximation \mythickbarredsquare{blue}{blue}. Right: pruning ratio obtained when targeting the level $1512$ with fixed weight (\mythickbarredsquare{red}{white}) and time-dependent weight obtained from brute force calculation (\mythickbarredsquare{blue}{white}) and the time-dependent weight obtained from the self-similarity approximation (\mythickbarredsquare{blue}{blue}). Plots correspond to the Lorenz 96 case with 64 degrees of freedom.}
\label{fig:postperf64}
\end{center}
\end{figure}

\begin{figure}[h]
\begin{center}
\includegraphics[width=0.30\textwidth]{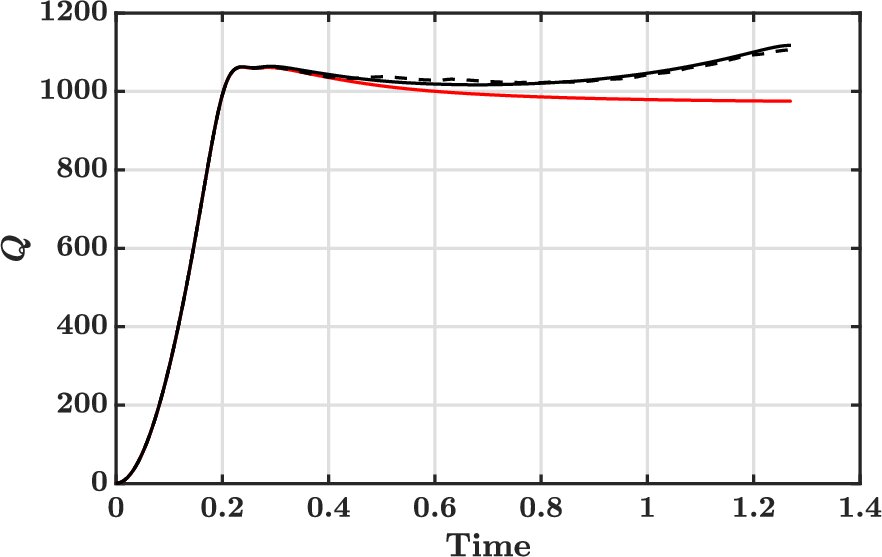}
\includegraphics[width=0.30\textwidth]{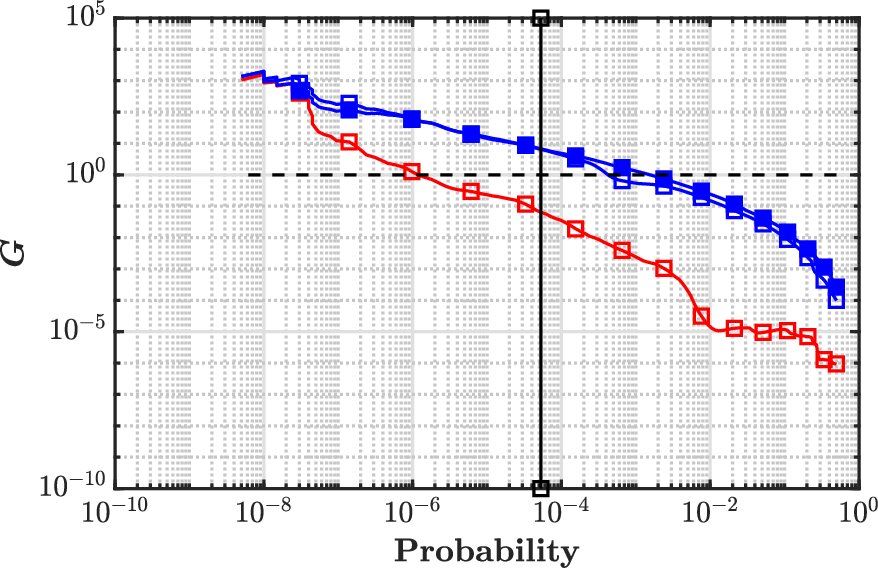}
\includegraphics[width=0.30\textwidth]{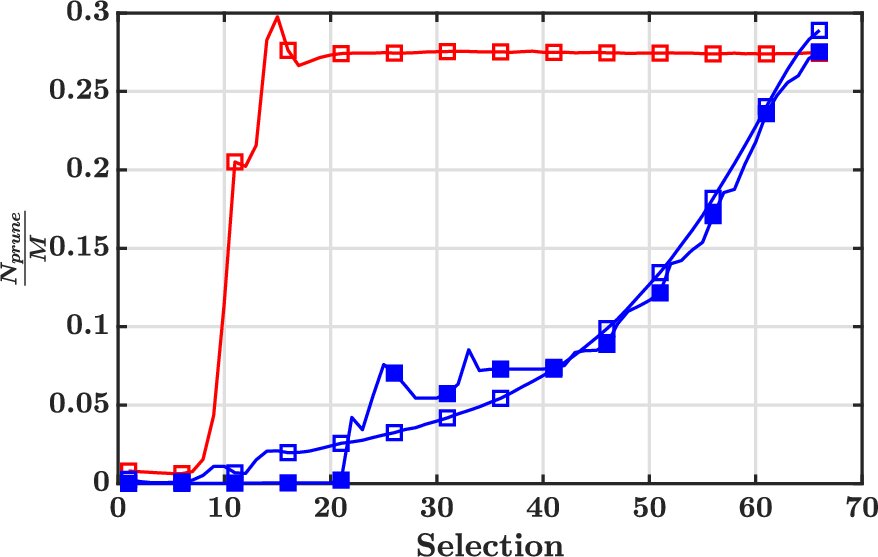}
\caption{Left: rare mean path $R_{1109}$ obtained from brute force computation (\mythickline{black}), from the self-similarity approximation (\mythickdashedline{black}) and ensemble average time history of the observable (\mythickline{red}). Middle: computational gain computed as Eq.~\ref{eq:compGain} when targeting the level $1109$ with the fixed weight (\mythickbarredsquare{red}{white}), the time-dependent weight obtained from the brute force calculation (\mythickbarredsquare{blue}{white}), the time-dependent weight obtained from the self-similarity approximation \mythickbarredsquare{blue}{blue}. Right: pruning ratio obtained when targeting the level $1109$ with fixed weight (\mythickbarredsquare{red}{white}) and time-dependent weight obtained from brute force calculation (\mythickbarredsquare{blue}{white}) and the time-dependent weight obtained from the self-similarity approximation (\mythickbarredsquare{blue}{blue}). Plots correspond to the Lorenz 96 case with 1024 degrees of freedom.}
\label{fig:postperf1024}
\end{center}
\end{figure}

Finally, the self-similar model is tested a posteriori. Using a brute force calculation run with $2500$ realizations, the paths leading to QoI ranging from $1100$ to $1225$ separated by a step size of $25$ are obtained for the 64-dimensional case. Paths leading to QoI ranging from $990$ to $1050$ separated by a step size of $10$ are obtained for the 1024-dimensional case. The rare mean path exceeding the thresholds $1512$ (for the 64-dimensional case) and $1109$ (for the 1024-dimensional case) are then estimated, similar to the procedure outlined in Sec.~\ref{sec:posteriori}. Figure \ref{fig:postperf64} (64-dimensional case) and Fig.\ref{fig:postperf1024} (1024-dimensional case) show that the self-similarity approach provides a reasonable approximation of the rare mean path. The self-similarity procedure is able to provide almost the same computational efficiency as the a priori procedure, while consistently outperforming the fixed weight method.

\begin{figure}[h]
\begin{center}
\includegraphics[width=0.45\textwidth]{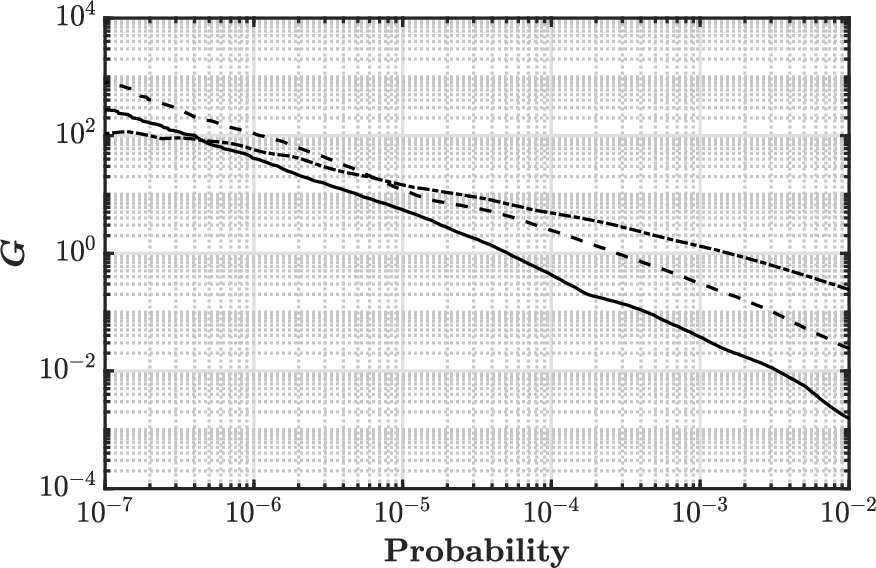}
\caption{Computational gain computed as Eq.~\ref{eq:compGain} when targeting the level located four standard deviations away from the mean for the 32-dimensional case (\mythickline{black}), the 64-dimensional case (\mythickdashedline{black}) and the 1024-dimensional case (\mythickdasheddottedline{black}).}
\label{fig:comparisonDimension}
\end{center}
\end{figure}

Finally, the performance of the ISP algorithm is compared between the 32, 64 and 1024-dimensional cases. For each case, the computational gain of the a posteriori analysis, obtained by targeting a level located four standard deviations away from the mean is plotted in Fig.~\ref{fig:comparisonDimension}. It can be seen that the performance does not degrade as the number of dimension increases. This observation suggests that the ISP method used here could be useful for very large-dimensional problems such as turbulent flows.

\bibliography{master}

\providecommand{\newblock}{}
\begin{thebibliography}{10}
\expandafter\ifx\csname url\endcsname\relax
  \def\url#1{{\tt #1}}\fi
\expandafter\ifx\csname urlprefix\endcsname\relax\def\urlprefix{URL }\fi
\providecommand{\eprint}[2][]{\url{#2}}

\bibitem{hassanaly2019computational}
Hassanaly M and Raman V 2019 {Computational Tools for Data-poor Problems in
  Turbulent Combustion} {\em AIAA Scitech 2019 Forum\/} p 0998

\bibitem{sornette1996stock}
Sornette D, Johansen A and Bouchaud J~P 1996 {\em Journal de Physique I\/} {\bf
  6} 167--175

\bibitem{weinan2007simplified}
Weinan E, Ren W and Vanden-Eijnden E 2007 {\em Journal of Chemical Physics\/}
  {\bf 126} 164103

\bibitem{solli2007optical}
Solli D, Ropers C, Koonath P and Jalali B 2007 {\em Nature\/} {\bf 450} 1054

\bibitem{ragone2018computation}
Ragone F, Wouters J and Bouchet F 2018 {\em Proceedings of the National Academy
  of Sciences\/} {\bf 115} 24--29

\bibitem{ghil2011extreme}
Ghil M, Yiou P, Hallegatte S, Malamud B, Naveau P, Soloviev A, Friederichs P,
  Keilis-Borok V, Kondrashov D, Kossobokov V {\em et~al.\/} 2011 {\em Nonlinear
  Processes in Geophysics\/} {\bf 18} 295--350

\bibitem{frei2001detection}
Frei C and Sch{\"a}r C 2001 {\em Journal of Climate\/} {\bf 14} 1568--1584

\bibitem{kim2013splitting}
Kim J, Bucklew J~A and Dobson I 2013 {\em IEEE Transactions on Power Systems\/}
  {\bf 28} 3010--3017

\bibitem{hassanaly2018ensemble}
Hassanaly M and Raman V 2019 {\em Proceeding of the Combustion Institute\/}
  {\bf 37}

\bibitem{siegmund1976importance}
Siegmund D 1976 {\em The Annals of Statistics\/}  673--684

\bibitem{morio2014survey}
Morio J, Balesdent M, Jacquemart D and Verg{\'e} C 2014 {\em Simulation
  Modelling Practice and Theory\/} {\bf 49} 287--304

\bibitem{de2005tutorial}
De~Boer P~T, Kroese D~P, Mannor S and Rubinstein R~Y 2005 {\em Annals of
  operations research\/} {\bf 134} 19--67

\bibitem{peherstorfer2016multifidelity}
Peherstorfer B, Cui T, Marzouk Y and Willcox K 2016 {\em Computer Methods in
  Applied Mechanics and Engineering\/} {\bf 300} 490--509

\bibitem{cerou}
C\'{e}rou D and Guyader A 2007 {\em Stochastic Analysis and Applications\/}
  {\bf 25}(2) 417--443

\bibitem{bouchet2019rare}
Bouchet F, Rolland J and Simonnet E 2019 {\em Physical review letters\/} {\bf
  122} 074502

\bibitem{teo2016adaptive}
Teo I, Mayne C~G, Schulten K and Leli\`{e}vre T 2016 {\em Journal of chemical
  theory and computation\/} {\bf 12} 2983--2989

\bibitem{laurie2015computation}
Laurie J and Bouchet F 2015 {\em New Journal of Physics\/} {\bf 17} 015009

\bibitem{lestang2018computing}
Lestang T, Ragone F, Br{\'e}hier C~E, Herbert C and Bouchet F 2018 {\em Journal
  of Statistical Mechanics: Theory and Experiment\/} {\bf 2018} 043213

\bibitem{wouters_bouchet}
Wouters J and Bouchet F 2016 {\em Journal of Physics A: Mathematical and
  Theoretical\/} {\bf 49} 374002

\bibitem{del2005genealogical}
Del~Moral P, Garnier J {\em et~al.\/} 2005 {\em The Annals of Applied
  Probability\/} {\bf 15} 2496--2534

\bibitem{dellago2002transition}
Dellago C, Bolhuis P~G and Geissler P~L 2002 {\em Advances in chemical
  physics\/} {\bf 123} 1--78

\bibitem{zhang2018rare}
Zhang B, Marzouk Y, Min B~Y and Sahai T 2018 {Rare Event Simulation of a
  Rotorcraft System} {\em 2018 AIAA Non-Deterministic Approaches Conference\/}
  p 1181

\bibitem{grafke2015instanton}
Grafke T, Grauer R and Sch{\"a}fer T 2015 {\em Journal of Physics A:
  Mathematical and Theoretical\/} {\bf 48} 333001

\bibitem{lorenz1996predictability}
Lorenz E~N 1996 {Predictability: A problem partly solved} {\em Proc. Seminar on
  predictability\/} vol~1

\bibitem{nemoto2016population}
Nemoto T, Bouchet F, Jack R~L and Lecomte V 2016 {\em Physical Review E\/} {\bf
  93} 062123

\bibitem{souza}
Souza A 2015 Instantons as a means to probe chaotic attractors Tech. rep. WHOI

\bibitem{kuramoto}
Kuramoto Y and Tsuzuki T 1976 {\em Progress of theoretical physics\/} {\bf 55}
  356--369

\bibitem{sivashinsky}
Sivashinsky G 1977 {\em Acta astronautica\/} {\bf 4} 1177--1206

\bibitem{etdrk4}
Kassam A~K and Trefethen L~N 2005 {\em SIAM Journal of Scientific Computing\/}
  {\bf 26} 1214--1233

\bibitem{gatechignition}
Sforzo B, H D, Wei S and Setzman J 2017 {\em Journal of Engineering for Gas
  Turbines and Power\/} {\bf 139} 031509

\bibitem{tang2019comprehensive}
Tang Y, Hassanaly M, Raman V, Sforzo B and Seitzman J 2019 {\em Combustion and
  Flame\/} {\bf 206} 158--176

\bibitem{malikmcs11}
Hassanaly M, Tang Y, Barwey S and Raman V 2019 {Analysis of the effect of
  turbulence on aircraft engine ignition} {\em Proceedings of the 11th
  Mediterranean Combustion Symposium\/}

\end{thebibliography}
\bibliographystyle{iopart-num}

\end{document}